\documentstyle[PASJadd]{PASJ95}
\draft 

\begin{document}
\setcounter{page}{1}

\title{{\it ASCA} Measurements of Metallicity and Temperature Distributions
in Three Clusters: A4059, MKW 3s and 2A 0335+096}

\author{ Ken'ichi {\sc Kikuchi}, Tae {\sc Furusho}, 
  Hajime {\sc Ezawa}, Noriko Y.\ {\sc Yamasaki}, and Takaya {\sc Ohashi} \\
{\it Department of Physics, Tokyo Metropolitan University,
1-1 Minami-Ohsawa,} \\
{\it  Hachioji, Tokyo 192-0397} \\
{\it E-mail(KK): kikuchi@phys.metro-u.ac.jp}
\\[6pt]
Yasushi {\sc Fukazawa} \\
{\it Department of Physics, University of Tokyo, 7-3-1 Hongo, Bunkyo-ku,  Tokyo
 113-0033} \\
and \\
Yasushi {\sc Ikebe} \\
{\it Max-Planck-Institut f\"{u}r Extraterrestrische Physik,
D-85748, Garching, Germany}}

\abst{We present {\it ASCA} results on the distributions of
  metallicity and temperature in 3 bright near-by clusters: A4059, MKW
  3s and 2A 0335+096.  A significant gradient in the metal abundance
  is detected in A4059, while other clusters suggest similar gradients
  with low significance.  These features together with recent results
  on AWM 7 and Perseus clusters suggest that metals injected in the
  ICM are not effectively mixed in the cluster space. Analysis of the
  GIS data, without explicitly including the cooling flow model, shows
  no substantial temperature drop at large radii (at half the virial
  radii) for the 3 systems. This is contrary to the recent results for
  30 clusters by Markevitch et al.\ (1998) who assume rather strong
  cooling flows.  The gas mass fraction of these clusters are 10--15\%
  within 1 Mpc, which suggests that baryonic fraction of about 20\%
  including the stellar mass is a common level in galaxy groups and
  clusters.}

\kword{Galaxies: abundances --- Galaxies: clustering --- Galaxies:
intergalactic medium --- X-Rays: spectra}

\maketitle

\thispagestyle{headings}

\section{Introduction}

Metallicity and temperature distributions in the hot intracluster
medium (ICM) in clusters of galaxies provide important information
about the history of cluster evolution and galaxy formation. The
heating of ICM is supposed to occur via gravitational energy release,
compression, and mergers over cosmological time scale.  Therefore,
temperature distribution reflects various physical
conditions in the ICM (strength of shocks, thermal conductivity,
magnetic field etc.)\ as well as cosmological parameters. Recent
observations, in particular from {\it ASCA} (Tanaka et al.\ 1994),
show non-radial and radial temperature structures in a number of
clusters. Among them, Markevitch et al.\ (1998) report that, when the
radius is scaled by the virial radius, the observed 30 clusters reveal
a very similar temperature profile.

Metals in the ICM are thought to be injected from individual galaxies. 
Proportionality between the mass of Fe in the ICM and the total
stellar mass in early-type galaxies provides a strong support to this
view (Tsuru 1992; Arnaud et al.\ 1992). However, it is still unclear
how metals have been injected into the cluster space. One scenario
predicts fast enrichment through galactic winds in the galaxy
formation period, and the other important process is a ram-pressure
stripping of the galaxy gas over a long time scale. Correct
determinations of the total amount of metals and its spatial
distribution in relation to those of galaxies and dark matter are useful
in determining the scenario of metal injection, which leads to
an understanding of the process of chemical evolution of galaxies.

This paper reports {\it ASCA} results on metallicity and temperature
distributions in 3 cD clusters: A4059, MKW 3s and 2A 0335+096. These
clusters are located at a distance of $z = 0.035 - 0.049$ and fit in
the field of view of the GIS. Their temperatures are lower than 4 keV,
therefore strong Fe emission line is expected. A4059 is associated
with a radio source PKS 2354-35 as a cD galaxy (Schwartz et al.\ 1991)
at $z = 0.0487$, showing an average ICM temperature $ kT \approx 3.5$
keV (David et al.\ 1993). {\it ROSAT} (Tr\"{u}mper 1983) has carried
out detailed observations of this cluster with PSPC and HRI (Huang and
Sarazin 1998). Although the HRI surface brightness profile indicates a
strong cooling flow with $\dot{M} \approx 184 M_\odot$ yr$^{-1}$, the
PSPC does not show the spectral signature of the cooling flow with
$\dot{M} < 80 M_\odot$ yr$^{-1}$.  The implied $\dot{M}$ is also
different from the PSPC deprojection results of $\dot{M} \approx 115
M_\odot$ yr$^{-1}$ by Allen and Fabian (1997). Temperature structure
is derived by Markevitch et al.\ (1998) based on the analysis assuming
a strong cooling flow. Preliminary {\it ASCA} results on this cluster
are given in Ohashi (1995), in which treatment of the energy dependent
point-spread function (PSF) is somewhat premature.

MKW 3s is a poor cluster at $z = 0.045$ with an average temperature
$kT \approx 3.7$ keV (David et al.\ 1993). Deprojection analysis of
{\it ROSAT} PSPC data shows a cooling flow rate of $\dot{M} \approx
161 M_\odot$ yr$^{-1}$ (Allen and Fabian 1997). Markevitch et al.\
(1998) report a possible temperature decline with radius based on {\it
ASCA} data.  2A 0335+096 is also a strong cooling flow cluster at $z =
0.035$ and $kT \approx 3.0$ keV (David et al.\ 1993). Irwin and
Sarazin (1995) report that the cooling flow rate takes a peak of
$\dot{M} = 400\ M_\odot$ yr$^{-1}$ at $r = 60$ kpc and remains the
same at larger radius. An extra absorbing matter associated with the
cluster is significantly seen inside a radius of 500 kpc. Sarazin et
al.\ (1995) reports high-resolution radio structures in the cluster
core.

These clusters are thought to be relaxed systems from their smooth
morphology and are useful in finding common properties of the ICM in
low temperature clusters. The circular structure enables a simple
analysis for the profiles, and the handling of stray light and
energy-dependent PSF of the {\it ASCA} X-ray telescope (XRT) is
relatively easy.  We investigate distributions of metallicity and
temperature in the three clusters of galaxies by taking into account
the complex properties of the XRT\@. We assume $H_0 = 50$ km s$^{-1}$
Mpc$^{-1}$ and $q_0 = 0.5$ throughout the paper, therefore an angular
separation of $1'$ corresponds to 78 kpc in A4059, 76 kpc in MKW 3s
and 59 kpc in 2A 0335+096. The 1 solar number abundance
of Fe relative to H is taken as $4.68 \times 10^{-5}$ (Anders \&
Grevesse 1989).

\section{{\it ASCA} Observations and Data Reduction}

Table 1 shows the log of {\it ASCA} observations.  We select the GIS
data with a cut-off rigidity $> 8$ GeV c$^{-1}$ and the telescope
elevation angle from the Earth rim $> 5^\circ$ (see Ohashi et al.\ 
1996 and Makishima et al.\ 1996 for the details of the GIS system).
Flare-like events due to the background fluctuation and data taken
with unstable attitude are discarded (Ikebe 1995).  The non X-ray and
the diffuse X-ray background is estimated from the archival data taken
during 1993--1994 (Ikebe 1995).  Since all targets were observed
during 1993--1994, the slow increase of the non X-ray background of
the GIS by $\sim 5$\% yr$^{-1}$ (Ishisaki et al.\ 1997, ASCA News No.\ 
4, 26) can be neglected.

Since the X-ray surface brightness of the 3 clusters drops below 50\%
of the background at $20'$ radius from the center, the stray light,
which are X-rays coming from outside of the field of view, can be
ignored.
This paper mainly presents GIS results, since it covers the larger 
region ($r>1$ Mpc from the center) and its detector response has less 
systematic errors than the SIS system.


\section{Radial Profile}

\subsection{{\it ROSAT} PSPC Data}

The PSF of the {\it ASCA} XRT has a half-power diameter of about $3'$
(Serlemitsos et al.\ 1995), which is insufficient to resolve the
intense central peaks (less than $\sim 1'$ which corresponds to 70--80
kpc) in these clusters. 
This situation is similar to that we found in Hydra-A cluster 
(Ikebe et al.\ 1997). 
We, therefore, use the {\it ROSAT} PSPC image as the model of the
surface brightness profile as has been done in Markevitch et al.\ 
(1996) and in Ikebe et al.\ (1997).

The {\it ROSAT} PSPC observed A4059, MKW 3s, and 2A 0335+096 on
November 21 in 1991, August 15 in 1992, and February 25 in 1991,
respectively.  We obtained the processed PSPC data from the archival
database provided by the {\it ROSAT} Scientific Data Center at Max
Planck Institut f\"{u}r Extraterrestrische Physik.  We exclude
contaminating sources in the PSPC data, and correct for the vignetting
using EXSAS software package (Zimmermann et al.\ 1991, EXSAS User's Guide).
Figure 1(a), (b) and (c) show the PSPC
image of 3 clusters.  Assuming the axial symmetry, we fit the
radial brightness profiles with a single $\beta$ model and a constant
background, but the model fails to represent the steep profile within $r
\ltsim 1'$.  Therefore we fit the data with a double $\beta$ model
represented by,
\begin{equation}
I(r) = I_1 \left[ 1+ \left(\frac{r}{R_{\rm c1}}\right)^2 \right]
        ^{-3\beta_1 +0.5} + I_2 \left[ 1+
        \left(\frac{r}{R_{\rm c2}}\right)^2 \right] ^{-3\beta_2 +0.5} +
        {\rm constant\; background}
\end{equation}
Parameters with suffixes 1 and 2 describe the central narrow component
and the extended emission, respectively.  $I_1$ and $I_2$ represent
normalization of each component.  The data in the radial range of
$\sim 20'$ are fitted with this model, without convolving the model
with the PSF of the PSPC system (figure 2).  We regard that, because
of the good angular resolution ($\sim 25''$) of the {\it ROSAT} PSPC
(Briel et al.\ 1997, {\it ROSAT} User's Handbook), neglecting the
effect of PSF dose not cause a problem in making a template profile
for the GIS data which has a broader response by a factor of $\sim
10$. When the value of $\beta_1$ is varied as a free parameter, the
PSPC data cannot constrain it because of strong coupling with other
parameters. We find that a fixed value of $\beta_1$ between 1.0--3.0
gives equally good fit to the PSPC radial profiles for all 3 clusters. 
Based on this study, we decide to fix $\beta_1$ at 1.0 in the following
analysis.  The center of the X-ray emission is chosen so that the core
radius takes the smallest value, i.e.\ giving the narrowest central
profile for each cluster. This method essentially places the center at
the peak of the X-ray emission. As summarized in table 2, this model
gives a reasonable fit ($\chi^2_\nu \ltsim 1.3$) to all clusters. We
notice that the narrow component for 2A 0335+096 is extremely strong
(as indicated by a large $I_1/I_2$ value by a factor of $\sim 5$)
compared with other clusters.


\subsection{GIS Data}
The GIS radial profiles can be described by a single $\beta$ model
because of the poor angular resolution. As shown in the previous
section, the PSPC data requires a double $\beta$ profile in the energy
range below 2 keV\@. The need of the double $\beta$ profile even in
the higher energy range has been shown for Hydra-A cluster based on
the joint analysis of PSPC and GIS data (Ikebe et al.\
1997). Therefore, we assume a double $\beta$ model in fitting the GIS
data.

To check the consistency between the radial profiles of GIS and PSPC,
we fit the GIS data in the energy range 0.7--2 keV with the double
$\beta$ model. The background (non X-ray and cosmic diffuse) is
subtracted from the data, and contaminating point sources seen in the
PSPC image are masked out. The double $\beta$ profile, described by
the parameter values given in table 2, is convolved with the PSF of
the XRT-GIS system. In fitting the GIS data, we choose the center
position which minimizes the core radius. The fit is performed over a
radius of $0'$ to $20'$. For MKW 3s, we include in the model a
foreground soft X-ray emission from the North Polar Spur, which is
described in section 4.3.  By adjusting the parameters of the wide
$\beta$ component, $\beta_2$ and $R_{c2}$, within the errors of the
PSPC fit, an acceptable fit for the 0.7--2 keV GIS data is obtained
(table 2).  In the later analysis, we fix the parameters of the double
$\beta$ model at the best-fit GIS values as listed in table 2.

\section{Spectral Analysis for Individual Annular Regions}

\subsection{Analysis with an ``Isothermal'' Model}

Examples of the background-subtracted GIS spectra for 3 annular
regions are shown in figure 3. One can see a general trend that the
equivalent width of Fe-K line decreases from the center to outer
region, most clearly in A4059 data. However, this feature is partly
flattened due to a flux mixing effect of the XRT as discussed in Ezawa
et al.\ (1997). Because of the extended tail of the PSF of the XRT,
the data in each ring are contaminated by photons from near-by rings.
In particular, the bright central region causes significant effect to
outer rings. Figure 4 shows relative fraction of the contaminating
flux from the other region compared with the region's own emission for
each ring. The contamination due to the central flux is particularly
strong in 2A 0335+096, and is about 40\% of the detected flux in the
outer ring ($r = 7'-8'$) originates from the central $r < 2'$) region.
The mutual flux contamination between different regions depends on the
distributions of surface brightness, temperature and metallicity.
This means that a simultaneous fit of the energy spectra in the entire
regions in the cluster is the only way to obtain the correct
distributions of temperature and metallicity.

When distributions of temperature and metallicity are uniform within
the cluster, we can create the proper response function based only on
the surface brightness profile and work out the correct spectral
parameters (see Honda et al.\ 1996).  This method, i.e.\ assuming the
uniform spectrum in the calculation of the response function, can be
used to carry out the first-order estimation of temperatures in
individual regions. We can, at the same time, examine whether this
particular model with an isothermal ICM and uniform metallicity is
consistent with the data or not. In calculating the response function
for 3 clusters, the double $\beta$ profile determined by the GIS data
analysis (table 2) is taken as the template brightness profile.  This
analysis gives a conservative estimate of temperature variation, if
present, in the cluster in the sense that the variation amplitude is
suppressed (Honda et al.\ 1996).

Using the above-described response, we first carried out a joint fit for
the GIS and SIS spectra in the inner region ($r < 2'$) where the
scattering effect is small. The GIS spectra for the annular regions
are then fitted to look at the radial variation of the ICM properties. 
We use the Raymond-Smith (Raymond \& Smith 1977; hereafter R-S)
thermal models with XSPEC ver.\ 9. Temperature ($kT$) and metal
abundance with solar ratio ($Z$) are varied as free parameters.  The
models give acceptable fit in most of the regions with $\chi^2/\nu
\ltsim 1.3$. Results for individual clusters are described in the
following sections.


\subsection{A4059} 
The GIS and SIS pulse-height spectra for the central region ($r < 2'$)
of the cluster are jointly fitted first with a single temperature R-S
model. The interstellar absorption $N_{\rm H}$ is fixed at the
Galactic value of $1.11 \times 10^{20}$ cm$^{-2}$ (Stark et al.\
1992). We obtain $\chi^2 / \nu = 543 / 350$, with best-fit
parameters $kT = 4.0$ keV and metal abundance $Z = 0.58$ solar,
respectively. See Table 3 for the errors of the parameters. This fit
is poor and the $\chi^2$ value is not formally acceptable at the 90\%
confidence. Two-temperature model is tried next with separate
abundance and common absorption as free parameters. This gives a
better fit with $\chi^2 / \nu = 500 / 346$, and the parameter
values are listed in Table 3. 
Assuming a common absorption for the hot and the cool components, the
metal abundance of the hot component ($kT = 4.2$ keV) becomes 0.62
solar and the cool component (0.9 keV) requires very low metallicity.
Hwang et al.\ (1997) showed that the abundance determined by Fe-L and
Fe-K lines are almost consistent for a range of plasma temperatures
$\sim 2-4$ keV, and suggested that the systematic offset between Fe-L
and Fe-K abundances was found if one did not consider excess
absorption for the cool component.  
If we carry out the fit with the same two-temperature model with
common metal abundance and separate absorption as free parameters, we
obtain $\chi^2 / \nu = 498 / 346$ with the common abundance
$0.60^{+0.09}_{-0.08}$ solar, and temperature $4.1^{+0.3}_{-0.2}$ keV
and $0.7^{+0.2}_{-0.1}$ keV for the hot and cool components,
respectively.  The absorption $N_{\rm H}$, assuming $z=0$ for the
absorber, is $7.9^{+0.5}_{-2.0} \times 10^{21}$ cm$^{-2}$ and
$8.1^{+3.3}_{-4.0} \times 10^{20}$ cm$^{-2}$ for the cool and hot
components, respectively.
Considering the systematic errors of 2--3 $\times 10^{20}$ cm$^{-2}$
on the $N_{\rm H}$ due to calibration uncertainty (Dotani et al.\ 
1996, ASCA News No.\ 4, 3), the absorption for the hot component is
almost consistent with the Galactic value.

The annular GIS spectra are then fitted with single temperature R-S
models with the $N_{\rm H}$ fixed at the Galactic value. Radial
distributions of temperature and metal abundance are shown in figure
5(a-1), (a-2) and table 4.  The temperature is almost constant with a
peak-to-peak fluctuation less than 0.6 keV. The metal abundance, on
the other hand, shows a systematic drop as a function of radius. The
central value of $0.75^{+0.14}_{-0.13}$ solar is very high for a
cluster, and it drops by a factor of about 3 at $r > 10'$. This
abundance profile is smoother than the actual one because of the PSF
effect as mentioned earlier, however it enables us to test whether the
abundance variation is significant or not because the correct response
for constant temperature and abundance is used in the fit. Taking the
7 values of the temperature and abundance for annular regions and
their $1\sigma$ confidence limits, we test the constant temperature or
constant abundance model with $\chi^2$ method. The fit gives
$\chi^2/\nu = 8.1/6$ for the temperature, and $21.1/6$ for the
abundance, respectively. Therefore, we can conclude that the abundance
variation in A4059 is significant at more than 99\% confidence, while
the temperature is consistent with a constant. If we replace the
abundance data in the innermost region with the combined GIS and SIS
result (0.62 solar), the abundance variation is still significant at
90\% confidence.

\subsection{MKW 3s} 
The spectrum in the central region is examined by jointly fitting the
GIS and SIS data within $r < 2'$ with R-S models. As shown in Table 3,
the single temperature model with fixed Galactic $N_{\rm H}$ gives a
poor fit with $\chi^2/\nu = 506/360$ and the two temperature model
gives a better fit with $\chi^2/\nu = 436/357$. Because of strong
coupling between spectral parameters, we need to assume a common metal
abundance for the hot and cool components in this fit.

The outer region of MKW 3s is significantly contaminated by soft X-ray
emission from the North Polar Spur (NPS). We fit the GIS spectrum in a
radial range of $12'-18'$ with thermal models because the emission
from the NPS is dominant in this region, and find that it can be
approximated by 0.28 keV thermal bremsstrahlung with a surface
brightness of $1.1 \times 10^{-14}$ erg cm$^{-2}$ s$^{-1}$
arcmin$^{-2}$ in 0.5 -- 2.0 keV\@.  This is close to the PSPC flux
($0.8 \times 10^{-14}$ erg cm$^{-2}$ s$^{-1}$ arcmin$^{-2}$) measured
in the same sky region with the {\it ROSAT} All Sky Survey (Snowden et
al.\ 1995).  The NPS flux in 0.5 -- 10 keV is about 30\% of the
cluster emission for the same solid angle at $r = 10'$ from the
cluster center.  Therefore, we include this component in fitting all
the data of MKW 3s.  Apart from this complexity in the foreground
emission, the cluster emission is described by a single temperature
R-S model in all regions. As indicated in figure 5(b-1), the
temperature does not vary significantly with the radius. The metal
abundance in this cluster shows a smaller gradient than in A4059. The
average value is about $0.35$ solar within $4'$, and it drops to 0.2 --
0.3 solar at $r > 8'$. Again, we test the constant
temperature/abundance models in the same way as for A4059, and the
$\chi^2$ fits show that no significant variation in temperature or
abundance is present in MKW3s at 90\% confidence.

\subsection{2A 0335+096} 

A strong cool component with $kT \sim 1$ keV and an excess $N_{\rm H}
\sim 1 \times 10^{21}$ cm$^{-2}$ are detected in the central region
($r < 3'$) of this cluster with the PSPC (Irwin and Sarazin 1995). A
single-temperature R-S fit to the combined GIS and SIS spectra in $r <
2'$ shows that the model is unacceptable with $\chi^2/\nu = 831/448$
(see Table 3). The single-temperature fit is unacceptable even if we
take the GIS data only.  A 2-temperature model with separate metal
abundance gives a better fit to with $\chi^2/\nu = 582/445$ for the
combined GIS and SIS data.  Both components show similar abundance
values as shown in Table 3. To better model the very strong cool
component, we include variable metal abundances (vR-S) for the cool
component. This model is used to avoid the problems arising from
calibration uncertainties and inappropriate modeling of atomic physics
(see Fabian et al.\ 1994) in the spectral fit.  Figure 6 shows the SIS
and GIS spectra fitted with the model. Free parameters in the fit are
2 normalizations, 2 temperatures, single $N_{\rm H}$ common to the 2
components, and 6 abundances (5 for the cool and 1 for the hot
component, respectively). The fit is acceptable with the best-fit
parameters as shown in table 5 with $\chi^2/\nu = 511/440$.  
The low Mg abundance is probably caused by incorrect modeling of 
Fe-L lines near the Mg-K lines (Fabian et al.\ 1994).
The $N_{\rm H}$ value of $3.0 \times 10^{21}$ cm$^{-2}$ is greater than
the Galactic one ($1.7 \times 10^{21}$ cm$^{-2}$, Stark et al.\ 1992),
which is consistent with the PSPC measurement by Irwin and Sarazin
(1995). When $N_{\rm H}$ is fixed at the Galactic value, GIS spectral
fit gives consistent spectral parameters as shown in table 5.

Assuming the two temperature component in all regions, we fit the GIS
spectra for the annular regions. A common $N_{\rm H}$ value for the
hot and cool components is fixed at the level of PSPC measurement
given in Irwin and Sarazin (1995). Also, temperature and abundance of
the cool component are fixed at those in table 5, so the spectral
shape of the cool component is not varied during the fit. This
treatment is necessary because GIS is rather insensitive to the
spectral features below $\sim 1$ keV\@. As a result, free parameters
in the fit are temperature, abundance and normalization of the hot
component, and normalization of the cool component. Spectra in the
outer regions ($r> 4'$) can be approximated by a single temperature
R-S model. In the outer regions, the single and two-temperature models
give the difference of only 10\% in the hot-component temperature.

The resultant values in Table 5 are tested against the constant model,
and we find that the abundance can be consistent at 90\% confidence
but that the temperature variation is significant at 99\%
confidence. It seems, however, likely that the 2 temperature model is
not good enough to model the strong complex emission at the center of
this cluster.


\section{Analysis with Image Response Matrix}

\subsection{Analysis Method}

In the previous section, we found a strong indication of the abundance
decline with radius in A4059. To evaluate the true values of
temperature and abundance, we need to jointly fit the spectra in the
annular regions using the image response matrix (Ikebe et al.\ 1997;
Markevitch et al.\ 1996).

The analysis with the ``isothermal'' model in the previous section
only gives an approximate solution of temperature and metal abundance.
When contribution of the stray light can be ignored, the combined fit
of the spectral data is performed by making so-called ``image response
matrix'' as described in detail in Ikebe et al.\ (1997) and in
Markevitch et al.\ (1996). Here, we will follow these approach and
evaluate true distributions of temperature and metal abundance.

In this analysis, the data format we deal with is a set of spatially
sorted spectra rather than energy sorted images. This preserves our
sensitivity to variations of temperature and metal abundance. The
spectral data of a cluster are jointly tested against a certain model
whose spatial distribution is described by a $\beta$-model. We
numerically construct 3-dimensional model clusters and predict GIS
spectra by convolving the sky image through image response
matrices. For all clusters, the GIS data are divided into 3 concentric
regions and are fitted simultaneously.

\subsection{A4059} 

Since the spectral fit for individual annular regions in the previous
section indicates that the ICM is isothermal, we will assume a
constant temperature for all regions and only look into the variations of
metal abundance.  In calculating a 3-dimensional model cluster, the
volume emissivity is described by,
\begin{equation} \epsilon(r) = n_{\rm gas}^2(r) \Lambda(T,Z(r)),
\end{equation}
where $\Lambda(T, Z(r))$ denotes the cooling function with metal
abundance $Z$. The hot gas distribution in the 3-dimensional space is
approximated as, 
\begin{equation} n_{\rm gas}^2(r) = n_{\rm 0}^2
\left[ \frac{I_1}{I_1+I_2} \left\{ 1 + \left(\frac{r}{R_{c1}}\right)^2
\right\}^{-3\beta_1} + \frac{I_2}{I_1+I_2} \left\{ 1 +
\left(\frac{r}{R_{c2}}\right)^2 \right\}^{-3\beta_2} \right],
\end{equation} 
where $n_{\rm 0}$ denotes central density of the hot gas, and
$I_1$, $I_2$, $R_{\rm c1}$, $R_{\rm c2}$, $\beta_1$, and $\beta_2$ are 
2-dimensional double $\beta$-model parameters in table 2.

A model with constant temperature and constant metal abundance is
examined in the first place.  For the 3 annular regions ($r = 0' -
2'$, $2' - 7'$ and $7' - 18'$), we fit the data in the energy range
0.7--10 keV with a single temperature R-S model.  Systematic error of
the image response matrix is included in $\chi^2$ value by adding 3\%
of the model flux (Ikebe et al.\ 1998).  This fit is acceptable with
the minimum $\chi^2$ value of 482.4 for $\nu = 452$, with $kT = 4.08
(3.97-4.19)$ keV and $Z = 0.42 (0.37-0.48)$ solar. As shown in
section 4.2 significant variation in the metal abundance exists in
this cluster, with the temperature consistent with a constant. The
purpose of the analysis in this section is to estimate the true radial
profile of the metal abundance based on the joint fit of the annular
spectra.

Using the image response matrix, we fit the 3 annular spectra by
varying the metal abundances in 3 regions separately. Free parameters
are the central density $n_0$, the common temperature $kT$, and 3
metal abundances for the 3 regions. The best fit $\chi^2$ value is
428.8 for $\nu = 450 $: a better fit than the constant abundance
model. Radial variation of the metal abundance is shown in figure
7(a), and the correlation of metal abundances in the 3 regions are
shown in figure 8, which show 90\% and 99\% confidence contours for
metal abundances for combinations of 2 different regions. The figure
shows that the abundances in the inner region and the outer region are
different with more than 99\% confidence.

\subsection{MKW 3s} 

Again, the ICM can be assumed to have a single temperature R-S
spectrum from the previous analysis. We assume different metal
abundances for 3 regions. A 0.28 keV thermal bremsstrahlung is added
for the North Polar Spur, and its normalization ($N_{\rm brems}$) is
varied in each region. Therefore, free parameters are the central
density $ n_{\rm 0,hot}$, the temperature $kT$, 3 metal abundances
($Z$), and 3 normalizations ($N_{\rm brems}$) for the 3 regions. The
abundance results are shown in figure 7(b). The best-fit points suggest
abundance gradient with $\chi^2 = 482.9$ with $\nu = 439$.
However, models with constant metal abundances for the energy range of
5.5--7.5 keV gives $\chi^2 = 19.6$ for $\nu = 13$, which is also
acceptable. Therefore, we cannot exclude the constant abundance model
for this cluster.

\subsection{2A 0335+096} 

The spectral fit for individual regions shows gradients in both
temperature and metallicity, so this is included in the combined
spectral fit.  We also found the strong cool component concentrated in
the central region. Since it is difficult to determine the spatial
distribution of the cool component with {\it ASCA} as mentioned
before, we assume that the distribution of the cool component is the
same as that of the narrow $\beta$-model,
\begin{equation}
        n_{\rm gas,cool}^2(r) = n_{\rm 0,cool}^2 \left[ 1 +
          \left(\frac{r}{R_{c1}}\right)^2 \right]^{-3 \beta_1}
\end{equation} 
with the parameters fixed at the values listed in table 2.  The volume
emissivity is given as,
\begin{equation}
        \epsilon(r) = f(r) \left[
          n_{\rm gas,hot}^2(r) \Lambda(T_{\rm hot}(r),Z_{\rm hot}(r))
          + n_{\rm gas,cool}^2(r) \Lambda(T_{\rm cool},Z_{\rm cool})
          \right],
\end{equation}
where, we assume that $n_{\rm gas,hot}^2$ has a double $\beta$
distribution as described in equation (3).
Since this model gives a radial profile which is a sum of a double
$\beta$ (hot) and a single $\beta$ (cool) models, we introduce a
correction factor $f(r)$ to adjust the emissivity profile to a double
$\beta$ distribution as seen in the PSPC data.  The actual value of
$f(r)$ is $0.6-1.1$ within $r<20'$.  The parameter $n_{\rm gas}$
differs from the true gas density, partly because of the correction
factor $f(r)$. Also, the present attempt in which the projected
emission is simply divided into the 2 independent hot and cool
components would inevitably result in a systematic deviation in the
estimated value of $n_{\rm gas}$ from the correct one.

Then we fit the 3 regions simultaneously, with free parameters $n_{\rm
0,hot}$, $n_{\rm 0,cool}$, 3 temperatures, and 3 abundances.  The
fixed parameters and their values are the same as those in the
individual spectral fit described in section 4.4. The results of the
spectral fits are shown in figure 7(c-1) and (c-2) for the temperature
of the hot component and the metal abundance, respectively.  The fit
is acceptable with $\chi^2 = 562.4$ with $\nu = 592$.  Constant metal
abundance gives an acceptable fit to the data in the energy range
5.5--7.5 keV, with $\chi^2 = 18.2$ for $\nu = 20$.  Therefore, we
cannot exclude the constant abundance model as well as in MKW 3s.
When we assume a step function with a radius of $2'$ for the surface
brightness distribution of the cool component, the temperature and
abundance of the hot component vary by less than 10\%.


\section{Mass Profile}

Based on the best fit model obtained in section 5, we calculate an
integrated mass distribution of the hot gas ($M_{\rm gas}$) and the
total gravitating mass ($M_{\rm total}$).  In calculating $M_{\rm
gas}$, we neglect the cool component and only include the hot
component since emission mechanism of the cool component very likely
involves different physics than the gravitational heating (see Fabian
1994, Makishima 1996, Ikebe et al.\ 1997). The double $\beta$ profile
with the parameters shown in table 2 with a constant temperature of
4.1 keV, 3.6 keV and 3.2 keV for A4059, MKW 3s and 2A 0335+096,
respectively, are assumed. The abundance gradient is approximated by a
smooth function and included for A4059. The results are shown in table
6 and in figure 9.

Spherical symmetry and hydrostatic equilibrium are assumed in
estimating $M_{\rm total}$. Again, presence of the cool component is
ignored. Since the constant temperature given above is also assumed,
the gravitating mass is given by $M_{\rm total} (r) = - (k T r / \mu G
m_{\rm H}) (d\ln n/d\ln r)$.  The same double $\beta$ profile for the hot
component is used as that in the calculation of $M_{\rm gas}$. As
shown in table 6 and in figure 9, the resultant mass profiles indicate
that the gas mass occupies 10 -- 15\% of the gravitating mass within 1
Mpc for all 3 clusters.


\section{Discussion}

{\it ASCA} observations have detected an abundance gradient in the
cluster of galaxies, A 4059. The feature is significant at 90\%
confidence when we take into account the effect of {\it ASCA} PSF\@.

Following the detection of metallicity gradient in clusters of
galaxies with Ginga and Einstein (White III et al.\ 1994), recent
studies show non-uniformity in the metallicity distribution in two
radial scales.  One is the small-scale ($\ltsim 100$ kpc) central
excess, often associated with the cool component around cD
galaxies. This feature is seen in Centaurus cluster (Ikebe et al.\
1998), Hydra A (Ikebe et al.\ 1997), Virgo cluster (Matsumoto et al.\
1996), A262 (David et al.\ 1996), and AWM 7 (Xu et al.\ 1997). Excess
metals are probably supplied by the cD galaxy and trapped in the deep
potential well.  We found that total iron mass within 100 kpc is about
$1.2 \times 10^{9} M_\odot$ for A4059.  Since the B band magnitude of
PKS 2354-35, cD galaxy of A4059, is 13.90 (Green et al.\ 1990), the
iron mass-to-light ratio around the cD is about $0.004
M_\odot/L_\odot$ for A4059.  This value is almost the same as the
average for clusters ($= 0.01-0.02 M_\odot/ L_\odot$, Arimoto et al.\
1997).  Therefore, the excess iron at the center of A4059 is within a
range which the cD galaxy can possibly supply.

The other finding is a large-scale gradient in the metal abundance
(over 500 kpc) which is now detected in 2 cD clusters; AWM 7 (Ezawa et
al.\ 1997) and Perseus cluster (Ezawa 1998).  It is likely that the
metals injected from individual galaxies are not effectively mixed in
the ICM and holds the original galaxy distribution, as indicated from
various models (e.g.\ Metzler and Evrard 1994). This feature would be
common to other clusters, but very few clusters have been studied from
{\it ASCA} with enough sensitivity to the outer region ($r > 500$
kpc). A rather uniform abundance distribution is found within central
200 kpc of the non-cD cluster A1060 (Tamura et al.\ 1996), which may
be different from the cD systems.

From the present result of A4059, we cannot definitely say whether the
high metallicity is only associated with the cD galaxy or it is
extended. Although the radial distribution of the
metal abundance suggests some large-scale gradient, more observations
with better statistics or better angular resolution is certainly
necessary. As shown in figure 7, 2A 0335+098 and MKW 3s also suggest
some gradient in the radial metallicity distribution, but we cannot
reject the case of spatially uniform abundance. Therefore, present
results have strengthened the possibility that the abundance gradient
is a common feature in many cD clusters.

Relation between cooling flows and metallicity gradient is discussed
by Fujita and Kodama (1995) and by Allen and Fabian (1998). Fujita and
Kodama show that cooling flows with a mass deposition rate $\dot{M}
\gtsim 100$ M$_\odot$ yr$^{-1}$ tends to make the abundance
distribution flatter because the flow compresses the distribution
pattern. This effect is probably compensated by the large metal supply
from the cD galaxy. Allen and Fabian (1998) shows that cooling flow
clusters have an average metallicity higher than that of non cooling
flow systems by a factor of $\sim 1.8$. They discuss that the central
region of high metallicity is not mixed by mergers in cooling flow
clusters. Based on the results on metallicity distribution so far
measured with {\it ASCA}, we consider it is highly likely that all the
cD clusters (usually reported to accompany strong cooling flows) have
a sharp concentration of metals in their centers. On the other hand,
the large scale gradient over $\sim 500$ kpc is probably unrelated to
either cD galaxy or cooling flows and rather common to all clusters
which have not experienced recent mergers.

One impact of the abundance gradient is its effect on the estimated
total mass of Fe ($M_{\rm Fe}$).  Because of the $n^2$ dependence of the
X-ray emissivity, a small amount of Fe concentrated in the center can
emit strong Fe-K line in the overall spectrum.  If we approximate the
abundance gradient in A4059 by a smooth exponential function with $Z =
1.0$ solar at the center declining to 0.20 solar at $r = 1$ Mpc, then
$M_{\rm Fe}$ is estimated to be $2.7 \times 10^{10} M_\odot$ in $r <
1$ Mpc.  
The total Fe mass inferred from the abundance gradient model is 
about 70 \% of that indicated if the abundance were uniform at
$Z \approx 0.45$ solar.

Markevitch et al.\ (1998) recently revealed temperature profiles 
in about 30 clusters. The temperature drops by a factor of 2
at half the virial radius.  Approximating the virial radius by
$r_{180}$, in which the mean mass density is 180 times the critical
density, and using the formula given by Markevitch et al.\ (1998) as
$r_{180} = 1.95 h^{-1} (T_{\rm x}/10 {\rm keV})^{1/2}$ Mpc, $r_{180}$
for the 3 clusters are 2.5 Mpc ($32'$) for A4059, 2.3 Mpc ($30'$) for
MKW 3s and 2.2 Mpc ($37'$) for 2A 0335+096. We note from
figure 5 and 7 that only A4059 shows a hint of temperature drop by less
than 20\% at $r = 0.5r_{180}$. Other clusters show no sign of
temperature drop, and 2A 0335+096 indicates a temperature rise in the
outer region. Part of this feature may be due to an exceptionally
strong cool component in the center. Even though 2 clusters, A4059 and
MKW 3s, are in the sample of Markevitch et al.\ (1998), the results are
significantly different.

The most likely cause of the difference is the treatment of cooling
flows. Markevitch et al.\ (1998) assumes cooling flow models in the
central region, while we simply fit the data with single temperature
models. Adding a cool component ($kT \sim 1$ keV) raises the
temperature of the ``hot'' component so that the average central
temperature is kept nearly constant. The GIS instrument is rather
insensitive to the cool component, so our analysis here is inevitably
simplified to some extent. On the other hand, Huang and Sarazin (1998)
note that the PSPC data of A4059 do not show spectral signature of the
cooling flow with $ \dot{M} < 80 M_\odot$ yr$^{-1}$ and discuss a
possibility that the gas may be ``young'' so that the cooling has not
yet established itself. Ikebe et al.\ (1997) show that the strong
central peak in the X-ray emission of Hydra-A cluster is dominated by
hot ($kT \sim 4$ keV) component rather than cool one, and discuss that
the cooling flow rate is 10 times less than the previous
estimation. These results indicate that a straightforward inclusion of
the cooling flow models tends to result in an overestimation of the
amount of cool gas (hence overestimation of the hot-gas temperature in
the center). We therefore suspect that the true temperature profile
would be somewhere between the present result and the Markevitch et
al.\ (1998) one. Close examination of temperatures in the very outer
regions of near-by clusters, which should be less affected by the
scattering effect of the XRT, would bring more definite conclusions
than the present sample. Also, joint analysis of the PSPC and ASCA
data in the central region would be interesting to know the nature of
the cool emission and the potential profile.

All three clusters indicate that the fraction of $M_{\rm gas}$ is
10--15\% of the total mass within $r < 1$ Mpc. It may be because these
clusters have the similar size and temperature. However, similar
values of the gas fraction have been reported in various clusters (see
Briel et al.\ 1992; Henry et al.\ 1993; Elbaz et al.\ 1995; David
1997).  This suggests that the baryonic fraction of about 20\%,
including the few \% stellar mass occupying a few \% of the
gravitational mass, is a common number in groups and clusters. The
constancy over a large range of cluster richness may suggest that it
is a cosmic average (therefore, $\Omega_B/\Omega \sim 0.2$) as
discussed by White et al.\ (1993). We need to point out, however, that 
clusters occupy only a few \% of the volume of the local universe and 
much of the space has so far been unexplored.

In the mass calculation we assume an isothermal temperature
distribution. If there is a radial drop of the temperature (in this
case a strong cool component has to be present in the cluster center),
the mass estimation at large radii ($r \sim 1$ Mpc) is affected in the
following manner. In the standard formula to describe the mass under a
hydrostatic equilibrium (e.g.\ Sarazin 1988, p.\ 178), the term $kTr$
is reduced to 70--80\% of the isothermal value based on the comparison
of Markevitch et al.\ (1998) value and the present ones at $r \sim 1$
Mpc. The term $d\ln T/d\ln r$ changes from 0 to about 0.5 if we assume
the average temperature profile in Markevitch et al.\ (1998). The two
terms give opposite effect on the total mass and it changes by only
about 10\%. Therefore, our mass estimation is unlikely to be seriously
wrong.

\vspace{1pc}\par

We would like to thank the {\it ASCA} Team for spacecraft operation
and data acquisition.  We also thank Dr.\ T.\ Tamura and Dr.\ S.\
Sasaki for important comments.  K. K. acknowledges support from the
Japan Society for the Promotion of Science for Young Scientists. This
work is partly supported by the Grants-in Aid of the Ministry of
Education, Science, Sports and Culture of Japan, 08404010.

%

\newpage
\section*{References}
\re Allen S.W., Fabian A.C.\ 1997, MNRAS 286, 583
\re Allen S.W., Fabian A.C.\ 1998, MNRAS 297, 63
\re Anders E., Grevesse N.\ 1989, Geochim. Cosmochim. Acta 53, 197
\re Arimoto N., Matsushita K., Ishimaru Y., Ohashi T., Renzini A.\ 1997, ApJ 477, 128
\re Arnaud M., Rothenflug R., Boulade O., Vigroux L., Vangioni-Flam E.\ 1992, A\&A 254, 49
\re Briel U.G., Henry J.P., B\"{o}hringer H.\ 1992, A\&A 259, 31
\re David L.P., Slyz A., Jones C., Forman W., Vrtilek S.D., Arnaud K.A.\ 1993, ApJ 412, 479
\re David L.P., Jones C., Forman W.\ 1996, ApJ 473, 692
\re David L.P.\ 1997, ApJL 484, 11
\re Elbaz D., Arnaud M., B\"{o}hringer H.\ 1995, A\&A 293, 337
\re Ezawa H., Fukazawa Y., Makishima K., Ohashi T., Takahara F., Xu H., Yamasaki N.Y.\ 1997, ApJ 490, L33
\re Ezawa H. 1998, Ph.D. thesis, Univ. of Tokyo
\re Fabian A.C., Arnaud K.A., Bautz M.W., Tawara Y.\ 1994, ApJL 436, 63.
\re Fabian A.C.\ 1994, ARA\&A 32, 227
\re Fujita Y., Kodama H.\ 1995, ApJ 452, 177
\re Green M.R., Godwin J.G., Peach J.V.\ 1990, MNRAS 243, 159
\re Henry J.P., Briel U.G., Nulsen P.E.J.\ 1993, A\&A 271, 413
\re Honda H., Hirayama M., Watanabe M., Kunieda H., Tawara Y., Yamashita K., Ohashi T., Hughes J.P., Henry J.P. 1996, ApJ 473, L71
\re Huang Z., Sarazin C.L.\ 1998, ApJ 496, 728
\re Hwang U., Mushotzky R.F., Loewenstein M., Markert T.H., Fukazawa, Y., Matsumoto, H.\ 1997, ApJ 476, 560 
\re Ikebe Y. 1995, Ph.D. thesis, Univ. of Tokyo
\re Ikebe Y., Makishima K., Ezawa H., Fukazawa Y., Hirayama M., Honda H., Ishisaki Y., Kikuchi K.\ et al.\ 1997, ApJ 481, 660
\re Ikebe Y., Makishima K., Fukazawa Y., Tamura T., Xu H., Ohashi T., Matsushita K.\ et al.\ 1998, ApJ submitted
\re Irwin J.A., Sarazin C.L.\ 1995, ApJ 455, 497
\re Makishima K., Tashiro M., Ebisawa K., Ezawa H., Fukazawa Y., Gunji S., Hirayama M., Idesawa E.\ et al.\ 1996, PASJ 48, 171
\re Makishima K.\ 1996, in X-Ray Imaging and Spectroscopy of Cosmic Hot Plasmas, ed F.\ Makino \& K.\ Mitsuda (Universal Academy Press, Tokyo) p137
\re Markevitch M., Mushotzky R., Inoue H., Yamashita K., Furuzawa A., Tawara Y.\ 1996, ApJ 456, 437
\re Markevitch M., Forman W.R., Sarazin C.L., Vikhlinin A.\ 1998, ApJ 503, 77
\re Matsumoto H., Koyama K., Awaki H., Tomida H., Tsuru T., Mushotzky R., Hatsukade I.\ 1996, PASJ 48, 201
\re Metzler C.A., Evrard A.E. 1994, ApJ 437, 564
\re Ohashi T.\ 1995, in Dark Matter, ed S.\ S.\ Holt \& C.\ L.\ Bennett (AIP, New York) p255
\re Ohashi T., Ebisawa K., Fukazawa Y., Hiyoshi K., Horii M., Ikebe Y., Ikeda H., Inoue H.\ et al.\ 1996, PASJ 48, 157
\re Raymond J.C., Smith B.W.\ 1977, ApJS 35, 419
\re Sarazin C.L., Baum S.A., O'Dea C.P.\ 1995, ApJ 451, 125
\re Schwartz D.A., Bradt H.V., Remillard R.A., Tuohy I.R.\ 1991, ApJ 376, 424
\re Serlemitsos P.J., Jalota L., Soong Y., Kunieda H., Tawara Y., Tsusaka Y., Suzuki H., Sakima Y.\ et al.\ 1995, PASJ 47, 105
\re Snowden S.L., Freyberg M.J., Plucinsky P.P., Schmitt J.H.M.M., Tr\"{u}mper J., Voges W., Edgar R.J., McCammon D., Sanders W.T.\ 1995, ApJ 454, 643
\re Stark A.A., Gammie C.F., Wilson R.W., Bally J., Linke R.A., Heiles C., Hurwitz M.\ 1992, ApJS 79, 77
\re Tamura T., Day C.S., Fukazawa Y., Hatsukade I., Ikebe Y., Makishima K., Mushotzky R.F., Ohashi T., Takenaka K., Yamashita K.\ 1996, PASJ 48, 671
\re Tanaka Y., Inoue H., Holt S.S.\ 1994, PASJ 46, L37
\re Tr\"{u}mper J.\ 1983, Adv.\ Space Res.\ 2, 241
\re Tsuru T. 1992, Ph.D. thesis, Univ. of Tokyo
\re White III R.E., Day C.S.R., Hatsukade I., Hughes J.P. 1994, ApJ 433, 583
\re White S.D.M., Navarro J.F., Evrard A.E., Frenk C.S. 1993, Nature 366, 429
\re Xu H., Ezawa H., Fukazawa Y., Kikuchi K., Makishima K., Ohashi T., Tamura T.\ 1997, PASJ 49, 9


\clearpage
\begin{table*}[tb]
  \begin{center}
    Table~1.\hspace{4pt}ASCA observation log of the 3 clusters\\
  \end{center}
  \begin{tabular*}{\textwidth}{@{\hspace{\tabcolsep}
    \extracolsep{\fill}}p{6pc}ccc}
    \hline\hline\\[-6pt]
      Target & Pointing (J2000.0)$^*$ & Observed Date (UT) &
        Exposure (sec)$^\dagger$ \\[4pt]\hline\\[-6pt]
      A4059      & $23^{\rm h}57^{\rm m}20^{\rm s}.4$, 
        $-34^{\rm d}43^{\rm m}10^{\rm s}$ & 1994/07/03 & 34151 \\
      MKW 3s      & $15^{\rm h}21^{\rm m}28^{\rm s}.9$, 
        $+07^{\rm d}42^{\rm m}36^{\rm s}$ & 1993/08/02 & 29258 \\
      2A 0335+096 & $03^{\rm h}38^{\rm m}45^{\rm s}.2$, 
        $+10^{\rm d}06^{\rm m}18^{\rm s}$ & 1994/08/22 & 40185 \\
    \hline
  \end{tabular*}
  \vspace{6pt}\par\noindent
  $*$ Center position of the GIS2 field of view.
  \par\noindent
  $\dagger$ Total exposure time of the GIS2 and 3 after the data screening
  described in section 2.
\end{table*}

\clearpage
\begin{table*}[tb]
  \begin{center}
    Table~2.\hspace{4pt}Result of fitting the radial profiles of PSPC
    and GIS with a double $\beta$ model.  Errors represent the 90\%
    confidence limit for one parameter.  For the GIS, parameters of
    the narrow component and normalization ratio ($I_1/I_2$) are fixed
    to the best-fit values derived by the PSPC.
  \end{center}
  \begin{tabular*}{\textwidth}{@{\hspace{\tabcolsep}
    \extracolsep{\fill}}rccccccc}
    \hline\hline\\[-6pt]
    & \multicolumn{2}{c}{Narrow component} & & \multicolumn{2}{c}{Extended component}\\ 
    \cline{2-3} \cline{5-6}\\
      \multicolumn{1}{l}{Target} & $\beta_1$ & $R_{\rm c1}$ & & $\beta_2$ & $R_{\rm c2}$ &  $I_1 / I_2$ & $\chi^2/\nu$ \\
      (Detector)                 &           & [arcmin]     & &           & [arcmin]     &  & \\
    \hline
      \multicolumn{2}{l}{A4059} \\
      (PSPC) & $1.00$  & $0.67$       & & $0.59$       & $1.48$       & 1.49         & 88.34/67 \\
             & (fixed) & (0.43--1.45) & & (0.57--0.67) & (1.23--3.28) & (0.25--2.29) &          \\
      (GIS)  & $1.00$  & $0.67$       & & $0.61$       & $1.53$       & 1.49         & 16.15/14 \\
             & (fixed) & (fixed)      & & (0.59--0.63) & (1.36--1.71) & (fixed)      &  \\
    \hline
      \multicolumn{2}{l}{MKW 3s} \\
      (PSPC) & $1.00$  & $0.76$       & & $0.61$       & $1.35$       & 1.43         & 76.69/74 \\
             & (fixed) & (0.51--1.02) & & (0.58--0.65) & (1.05--1.89) & (0.64--2.21) &          \\
      (GIS)  & $1.00$  & $0.76$       & & $0.63$       & $1.35$       & 1.43         & 12.59/10 \\
             & (fixed) & (fixed)      & & (0.59--0.67) & (1.06--1.57) & (fixed)      &  \\
    \hline
      \multicolumn{2}{l}{2A 0335+096} \\
      (PSPC) & $1.00$  & $0.93$       & & $0.64$       & $1.93$       & 8.10         & 88.34/74 \\
             & (fixed) & (0.88--0.97) & & (0.61--0.67) & (1.62--2.25) & (6.35--10.72) &          \\
      (GIS)  & $1.00$  & $0.93$       & & $0.64$       & $1.50$       & 8.10         & 17.07/14 \\
             & (fixed) & (fixed)      & & (0.62--0.66) & (1.27--1.64) & (fixed)      &  \\
    \hline
  \end{tabular*}
\end{table*}

\clearpage
\begin{table*}[tb]
  \begin{center}
    Table~3.\hspace{4pt}Result of the simultaneous spectral fitting of 
    GIS and SIS for the central $2'$ region. Errors represent the 90\% 
    confidence limit for one parameter.
  \end{center}
  \begin{tabular*}{\textwidth}{@{\hspace{\tabcolsep}
    \extracolsep{\fill}}lrccc}
    \hline\hline\\[-6pt]
    Target & & A4059  & MKW3s & 2A 0335+096 \\
    \hline
    Single R-S 
      & $kT$         & 4.0          & 3.7          & 2.7 \\
      & (keV)        & (3.8--4.1)   & (3.6--3.8)   & (2.6--2.8) \\
      & $Z$          & 0.58         & 0.49         & 0.71 \\
      & (solar)      & (0.50--0.66) & (0.43--0.56) & (0.66--0.77) \\
      & $\chi^2/\nu$ & 543.4/350    & 506.4/360    & 831.4/448 \\
    \hline
    Double R-S 
      & $kT_1$        & 0.9          & 1.4          & 1.3 \\
      & (keV)         & (0.8--1.2)   & (1.0--2.0)   & (1.2--1.4) \\
      & $Z_1$         & 0.04         & 0.42         & 0.63 \\
      & (solar)       & ($<0.12$)    & (0.35--0.49) & (0.46--1.01) \\
      & $kT_2$        & 4.2          & 3.9          & 3.3 \\
      & (keV)         & (3.9--4.9)   & (3.0--5.8)   & (3.0--3.7) \\
      & $Z_2$         & 0.62         & 0.42         & 0.55 \\
      & (solar)       & (0.53--0.67) & (fixed to $Z_1$) &(0.47--0.64) \\
      & $\chi^2/\nu$  & 500.0/346    & 436.6/357    & 581.8/444 \\
  \hline
  \end{tabular*}
\end{table*}

\clearpage
\begin{table*}[tb]
  \begin{center}
    Table~4.\hspace{4pt}Spectral fitting results with ``Isothermal''
    models.  Errors represent the 90\% confidence limit for one
    parameter.
  \end{center}
  \begin{tabular*}{\textwidth}{@{\hspace{\tabcolsep}
    \extracolsep{\fill}}lrccc}
    \hline\hline\\[-6pt]
      Target & radius  & $kT$ (keV)          & $Z$ (solar) & $\chi^2/\nu$ \\
    \hline
      A4059       & $0'-2'$   & 3.83 (3.68--3.99) & 0.75 (0.62--0.89) & 134.8/129 \\
                  & $2'-4'$   & 4.14 (4.00--4.30) & 0.45 (0.35--0.53) & 173.5/129 \\
                  & $4'-6'$   & 4.07 (3.88--4.27) & 0.40 (0.30--0.51) & 73.0/62 \\
                  & $6'-8'$   & 3.91 (3.67--4.16) & 0.48 (0.33--0.64) & 86.0/62 \\
                  & $8'-10'$  & 3.97 (3.63--4.36) & 0.25 (0.08--0.44) & 25.1/31 \\
                  & $10'-14'$ & 3.83 (3.44--4.31) & 0.22 (0.02--0.45) & 35.2/30 \\
                  & $14'-18'$ & 3.51 (2.97--4.30) & 0.71 (0.20--1.47) & 32.0/30 \\
    \hline
      MKW 3s       & $0'-2'$   & 3.59 (3.42--3.77) & 0.31 (0.22--0.41) & 142.0/128 \\
                  & $2'-4'$   & 3.56 (3.42--3.72) & 0.38 (0.29--0.48) & 140.5/128 \\
                  & $4'-6'$   & 3.44 (3.27--3.65) & 0.34 (0.23--0.46) & 56.7/61 \\
                  & $6'-8'$   & 3.37 (3.10--3.67) & 0.17 (0.04--0.31) & 53.0/61 \\
                  & $8'-12'$  & 3.41 (3.09--3.79) & 0.21 (0.05--0.40) & 30.0/30 \\
                  & $12'-18'$ & 3.65 (3.06--4.49) & 0.13 (0.00--0.44) & 34.0/30 \\
    \hline
      2A 0335+096 & $0'-2'$   & 3.38 (3.28--3.48) & 0.52 (0.44--0.62) & 159.0/126 \\
                  & $2'-4'$   & 3.26 (3.09--3.45) & 0.42 (0.35--0.50) & 189.4/128 \\
                  & $4'-6'$   & 3.51 (3.26--3.81) & 0.37 (0.29--0.46) & 122.7/128 \\
                  & $6'-8'$   & 3.08 (2.97--3.30) & 0.49 (0.38--0.60) & 65.4/61 \\
                  & $8'-10'$  & 3.63 (3.14--4.43) & 0.35 (0.19--0.52) & 49.5/61 \\
                  & $10'-14'$ & 4.73 (4.07--5.81) & 0.33 (0.18--0.48) & 25.3/29 \\
                  & $14'-18'$ & 4.79 (4.09--6.70) & 0.21 (0.00--0.43) & 34.2/29 \\
    \hline
  \end{tabular*}
  \vspace{6pt}\par\noindent

  NOTE: Interstellar absorption ($N_{\rm H}$) is fixed to the Galactic
  value, $1.11 \times 10^{20}$ cm$^{-2}$ for A4059, and $2.88 \times
  10^{20}$ cm$^{-2}$ for MKW 3s.  For 2A 0335+096, we fixed $N_{\rm
    H}$ at the sum of the Galactic and the excess absorption measured
  by the PSPC (Irwin \& Sarazin 1995): $3.03 \times 10^{21}$
  cm$^{-2}$, $2.40 \times 10^{21}$ cm$^{-2}$, $2.20 \times 10^{21}$
  cm$^{-2}$ and $1.72 \times 10^{21}$ cm$^{-2}$ for the region of
  $0'-2', 2'-4', 4'-8'$ and $8'-18'$, respectively.
\end{table*}

\clearpage
\begin{table*}[tb]
  \begin{center}
    Table~5.\hspace{4pt}Result of the spectral fit for the central $2'$
    region of 2A 0335+096 with a 2 temperature model.
  \end{center}
  \begin{tabular*}{\textwidth}{@{\hspace{\tabcolsep}
    \extracolsep{\fill}}rl}
    \hline\hline\\[-6pt]
      \multicolumn{1}{l}{Cool component (vR-S)} \\
        Temperature (keV)                  & $1.42 (1.35-1.65)$ \\
        Abundance (solar)                  &  \\
        \multicolumn{1}{r}{He, C, N}       & $1.0$ (fixed) \\
        \multicolumn{1}{r}{O, Ne}          & $1.36 (0.28-2.91)$ \\
        \multicolumn{1}{r}{Mg}             & $0.00 (<0.09)$ \\
        \multicolumn{1}{r}{Si}             & $0.91 (0.64-1.58)$ \\
        \multicolumn{1}{r}{S, Ar, Ca}      & $0.27 (0.03-0.48)$ \\
        \multicolumn{1}{r}{Fe, Ni}         & $0.69 (0.52-1.12)$ \\
    \hline
      \multicolumn{1}{l}{Hot component (R-S)} \\
        Temperature (keV)                  & $3.48 (3.10-4.27)$ \\
        Abundance (solar)                  & $0.54 (0.46-0.65)$ \\
    \hline
        $N_{\rm H}$ (cm $\times 10^{21}$)  & $3.03 (2.79-3.27)$ \\
        $\chi^2/\nu$                       & $510.7/440$ \\
    \hline
  \end{tabular*}
\end{table*}

\clearpage
\begin{table*}[tb]
  \begin{center}
    Table~6.\hspace{4pt} Integrated mass of the hot gas and the total
    gravitating matter.
  \end{center}
  \begin{tabular*}{\textwidth}{@{\hspace{\tabcolsep}
    \extracolsep{\fill}}p{6pc}ccc}
    \hline\hline\\[-6pt]
      Target & Average temperature & Gas mass ($M_{\rm gas}$)  & Gravitating mass ($M_{\rm total}$) \\
             & [keV]              & ($<1$ Mpc) [$M_{\odot}$]  & ($<1$ Mpc) [$M_{\odot}$]  \\
    \hline
      A4059      & 4.1            & $3.3\times10^{13}$        & $2.6\times10^{14}$ \\
      MKW 3s      & 3.6            & $3.4\times10^{13}$        & $2.4\times10^{14}$ \\
      2A 0335+096 & 3.2            & $3.1\times10^{13}$        & $2.1\times10^{14}$ \\
    \hline
  \end{tabular*}
\end{table*}

\clearpage
\begin{fv}{1}{10cm}{Contour plot of the PSPC image in the 0.4--2.0 keV band
    for (a) A4059, (b) MKW 3s and (c) 2A 0335+096.  These images are
    convolved with a Gaussian of $0'.5$ FWHM.  The contour levels are
    logarithmic scale with a step size of factor of 2.  The positions
    excluded from the analysis due to contaminating sources are
    indicated by triangles.}
\end{fv}

\bigskip
\begin{fv}{2}{10cm}{PSPC radial profiles fitted with the double $\beta$ models.
    (a) A4059, (b) MKW 3s and (c) 2A 0335+096.  The data (open
    circles) gives the X-ray surface brightness in the energy band of
    0.7--2 keV.  Dashed curves show narrow $\beta$ component, dotted
    curves show extended one, and the constant background is indicated
    by the solid line.  }
\end{fv}

\begin{fv}{3}{10cm}{Background-subtracted energy spectra for 3 
    concentric annular regions obtained with GIS.  (a) A4059, (b) MKW
    3s and (c) 2A 0335+096.  Each cluster shows a general trend that
    the equivalent width of Fe K line ($\sim$ 6--7 keV) decreases from
    the center to outer region.  }
%
%
\end{fv}

\begin{fv}{4}{10cm}{Simulation result of the contaminating flux from the 
annular sky regions to individual regions of the GIS for
(a) A4059, (b) MKW 3s and (c) 2A0335+096.
The horizontal axis shows the radius ranges of the GIS, and vertical
axis shows the fraction of photons originated in different sky regions.
We assume that the intensity profile of each cluster is represented
by double $\beta$ model with the parameter values shown in Table 3.
}
\end{fv}

\begin{fv}{5}{}{Results from the GIS spectral fit with the ``Isothermal''
    model: (a-1) temperature of A4059, (a-2) abundance of A4059, (b-1)
    temperature of MKW 3s, (b-2) abundance of MKW 3s, (c-1) temperature of
    2A0335+096 and (c-2) abundance of 2A0335+096.  Error bars represent the
    90\% confidence level for a single interesting parameter.
    Diamonds show results for broad binning in the radial direction.}
\end{fv}

\bigskip
\begin{fv}{6}{10cm}{The top panel shows the SIS and GIS spectrum of 2A 0335+096
    in the central $2'$ region, and the bottom one indicates the
    deviation of the data in the form of $\chi$.  Crosses represents
    the data, and the solid line shows the 2 temperature model
    summarized in Table 4.}
\end{fv}

\begin{fv}{7}{10cm}{Results from the  GIS spectral fit with the image 
    response matrix method described in section 4.  The 4 panels show
    (a) abundance of A4059, (b) abundance of MKW 3s, (c-1) temperature
    of 2A 0335+096 and (c-2) abundance of 2A 0335+096, respectively.
    Error bars are the 90\% confidence level for a single interesting
    parameter. A Smooth exponential curve is fitted to the results in
    (a).

}
\end{fv}

\begin{fv}{8}{10cm}
  {Confidence contour for the metal abundances in 2 different annular
    regions in A4059.  The inner and the outer contours correspond to
    90\% and 99\% limits for 2 parameters.  The solid straight line
    indicates the case for the same abundance.}
\end{fv}

\begin{fv}{9}{10cm}{
    Integrated mass profile of ICM ($M_{\rm gas}$) and total
    gravitating mass ($M_{\rm total}$) as a function of distance from
    the center for (a) A4059, (b) MKW 3s and (c) 2A 0335+096.  The gas
    fraction ($M_{\rm gas} / M_{\rm total}$) is also indicated in the
    bottom panels.}
\end{fv}

\label{last}
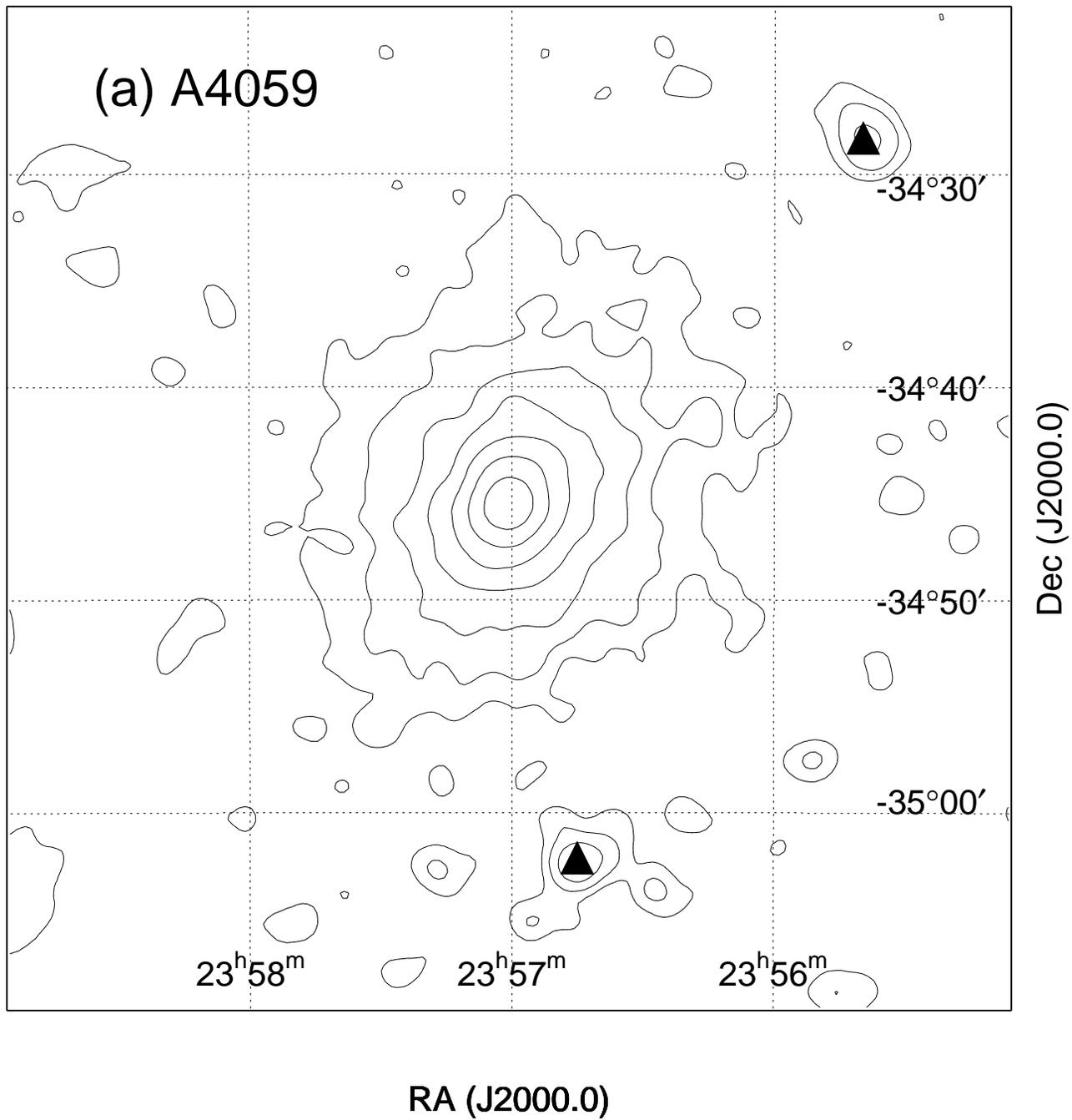
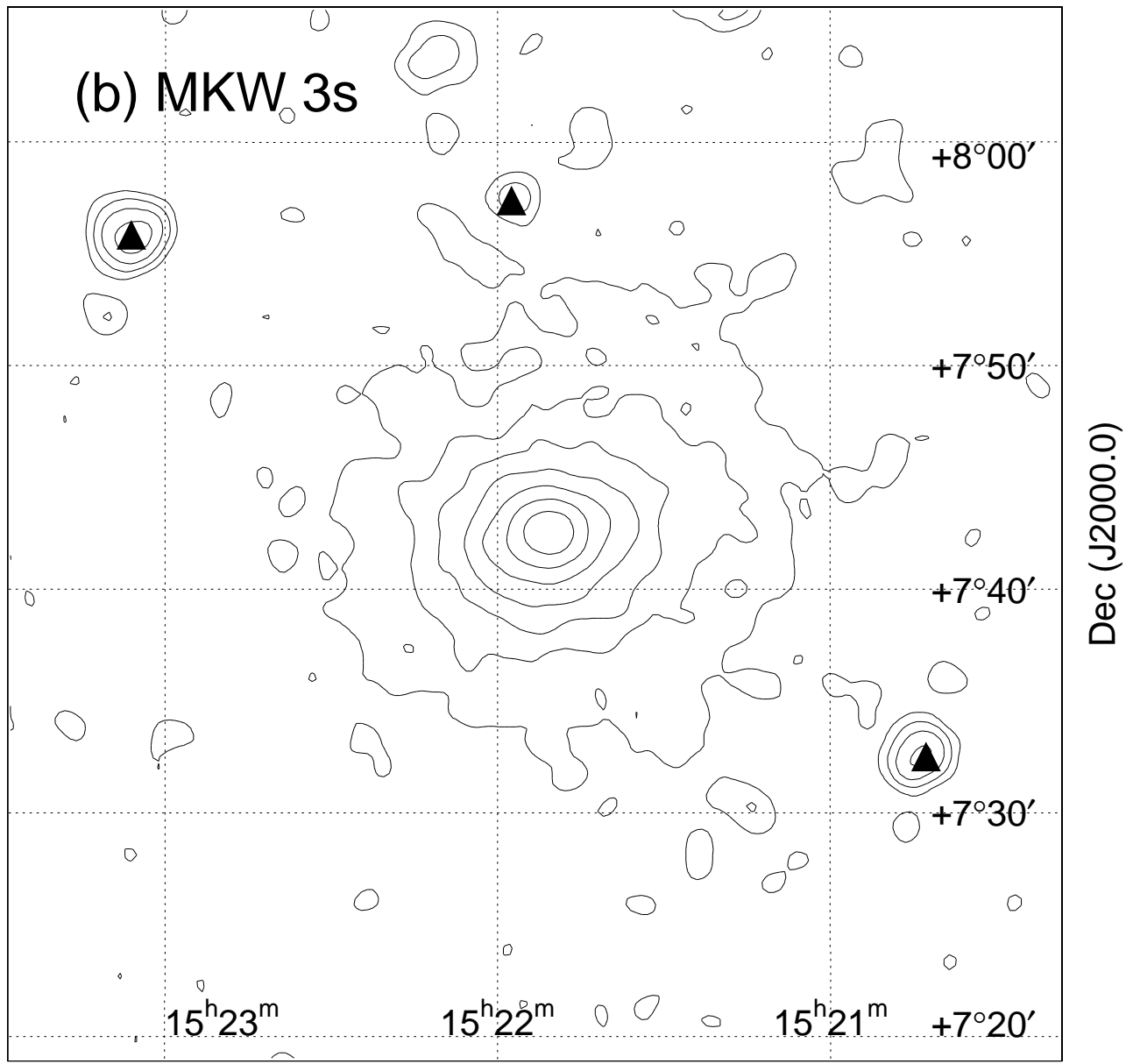
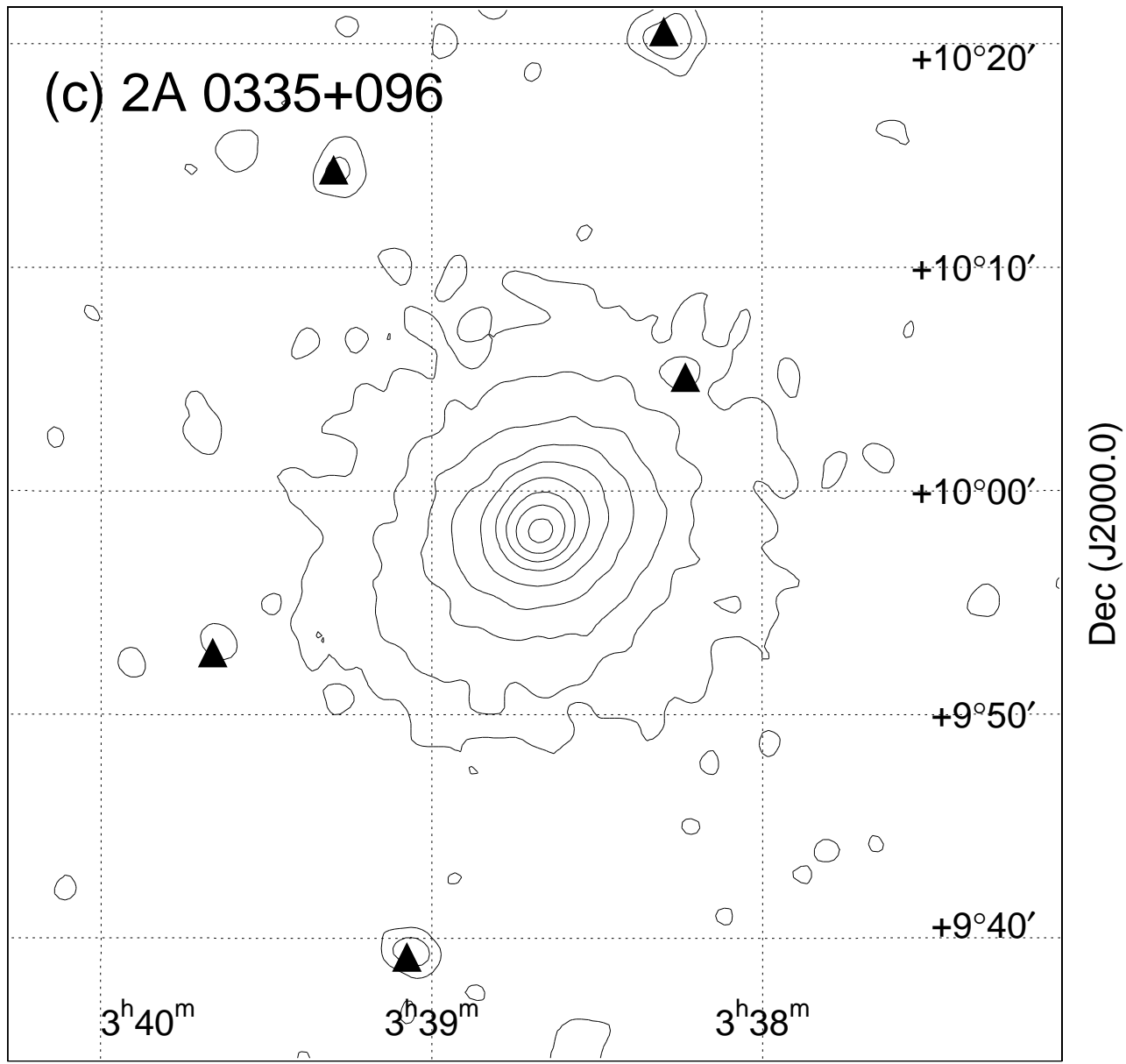
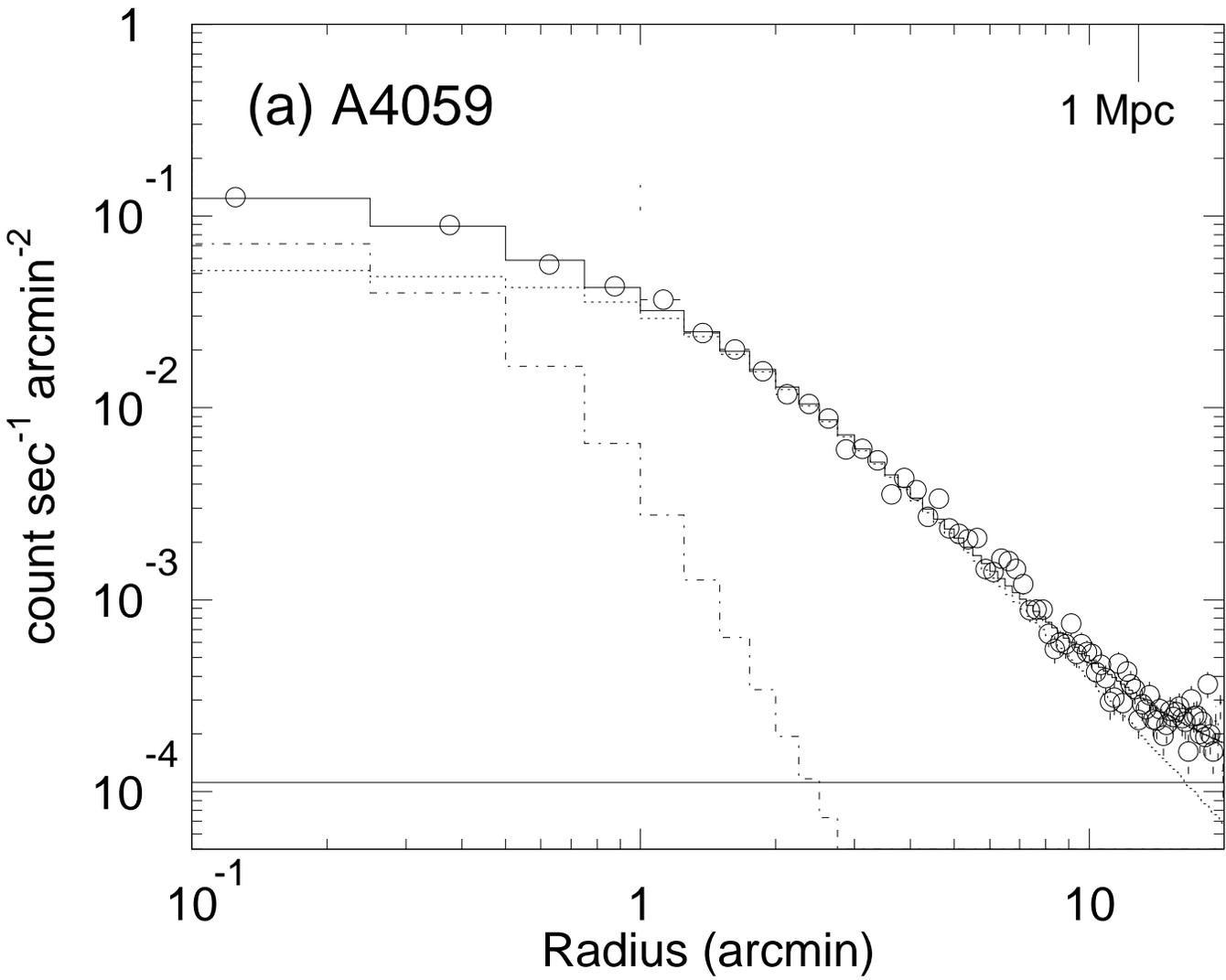
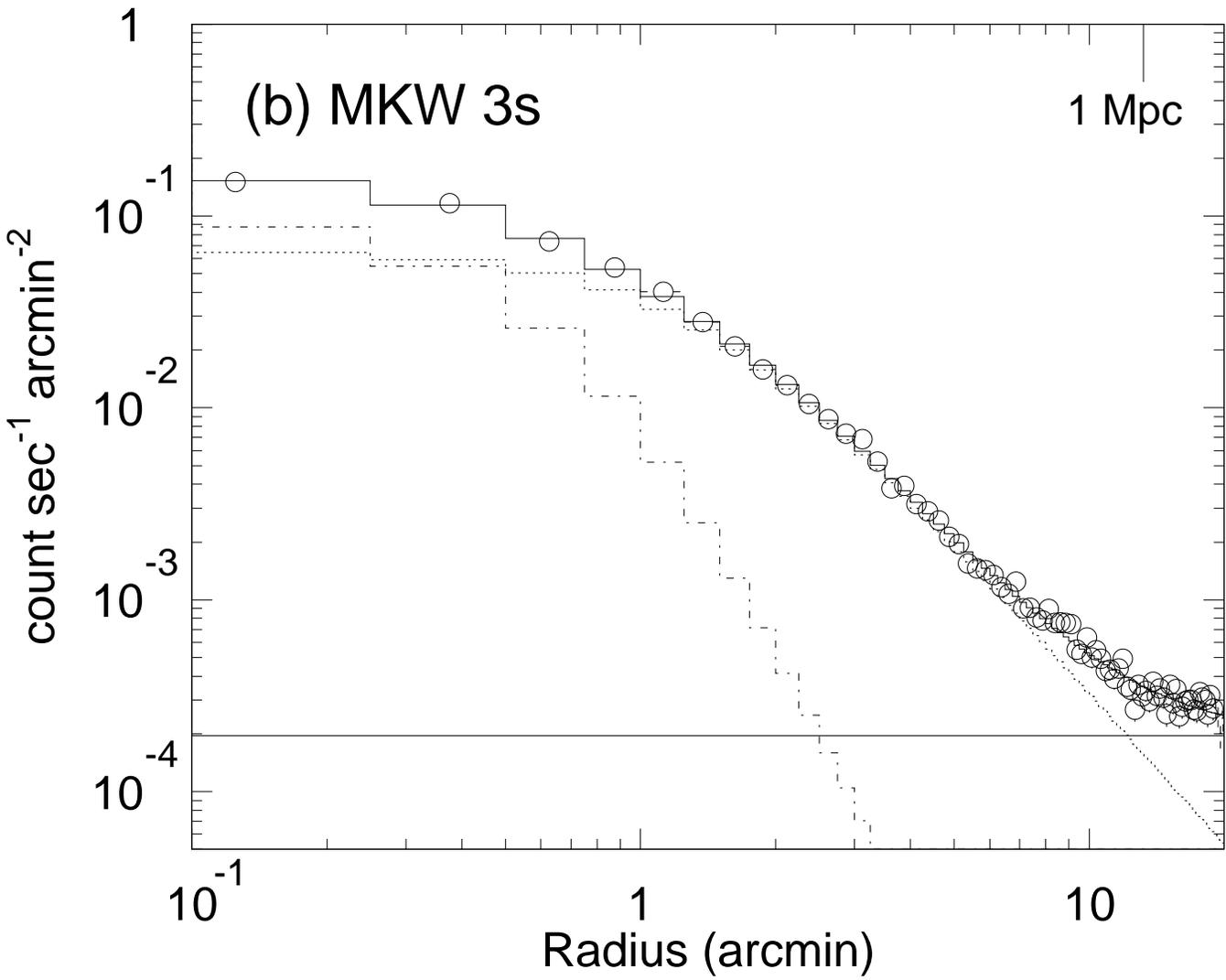
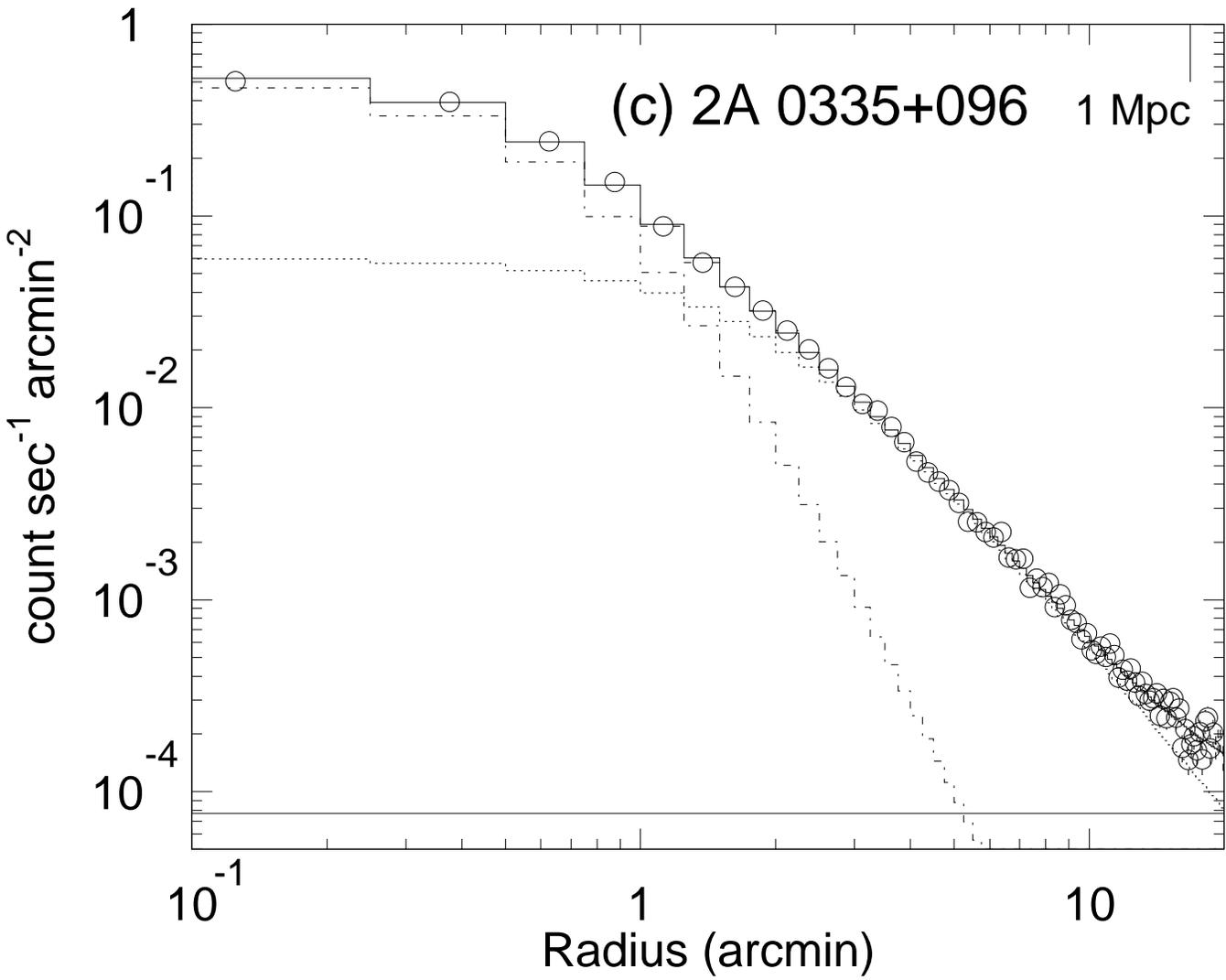
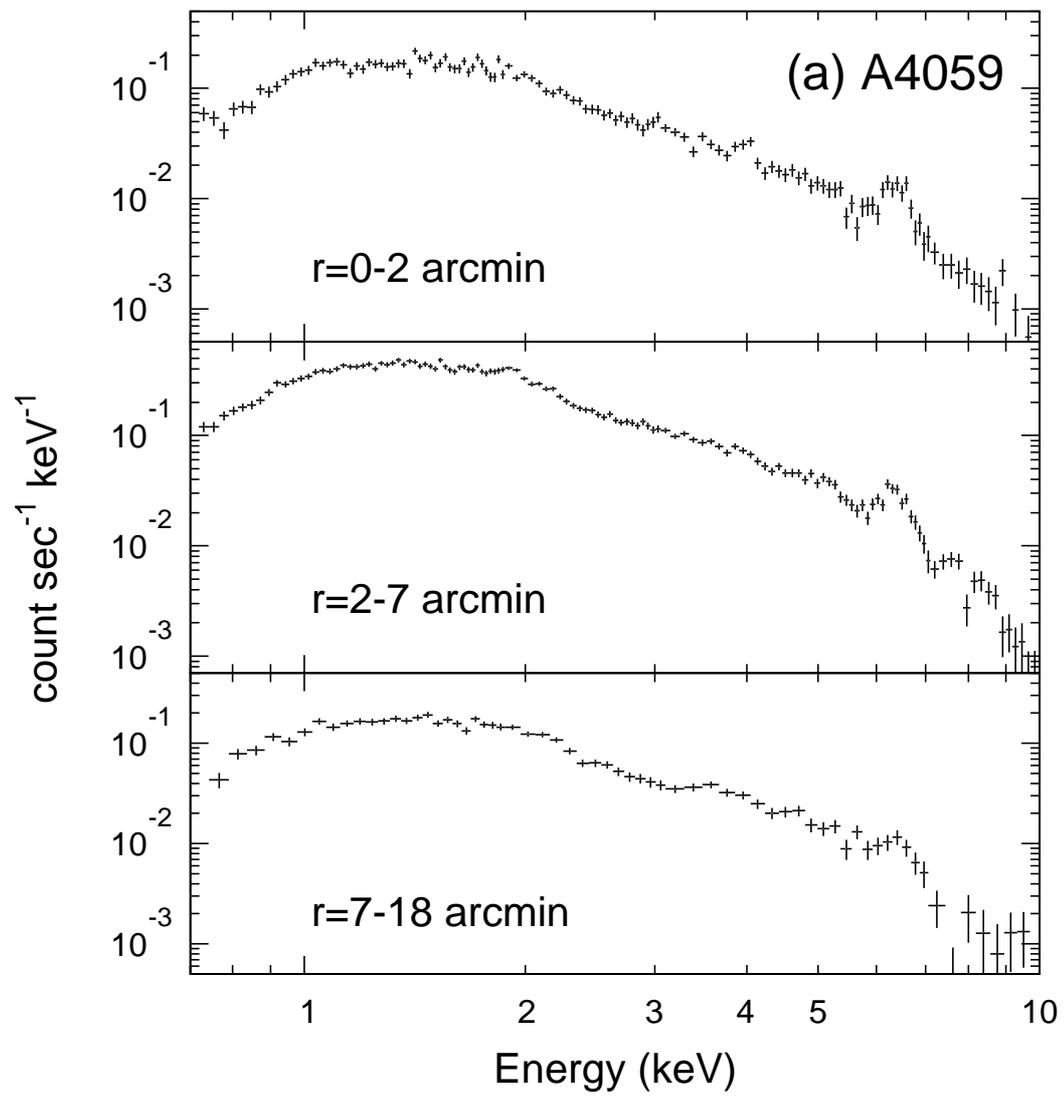
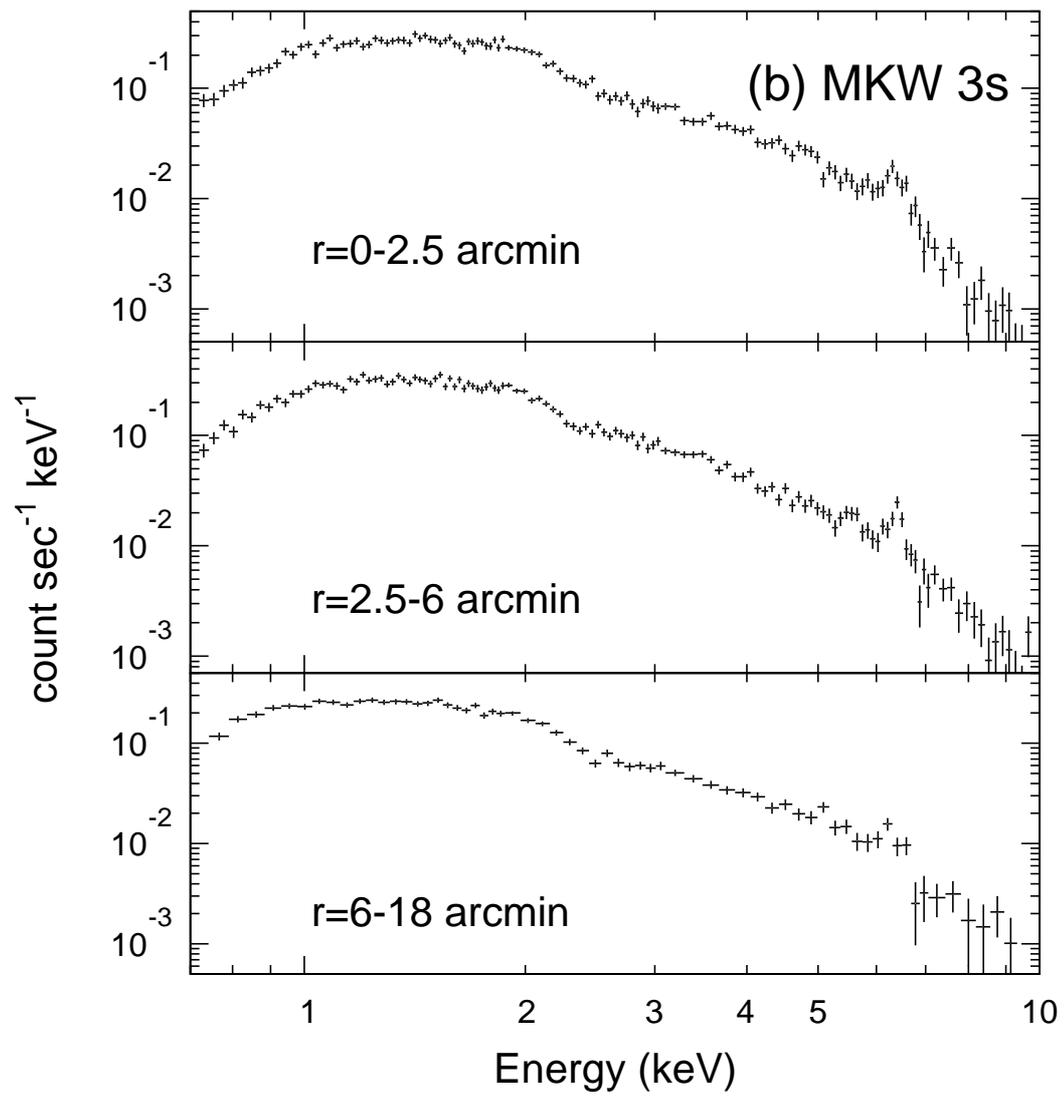
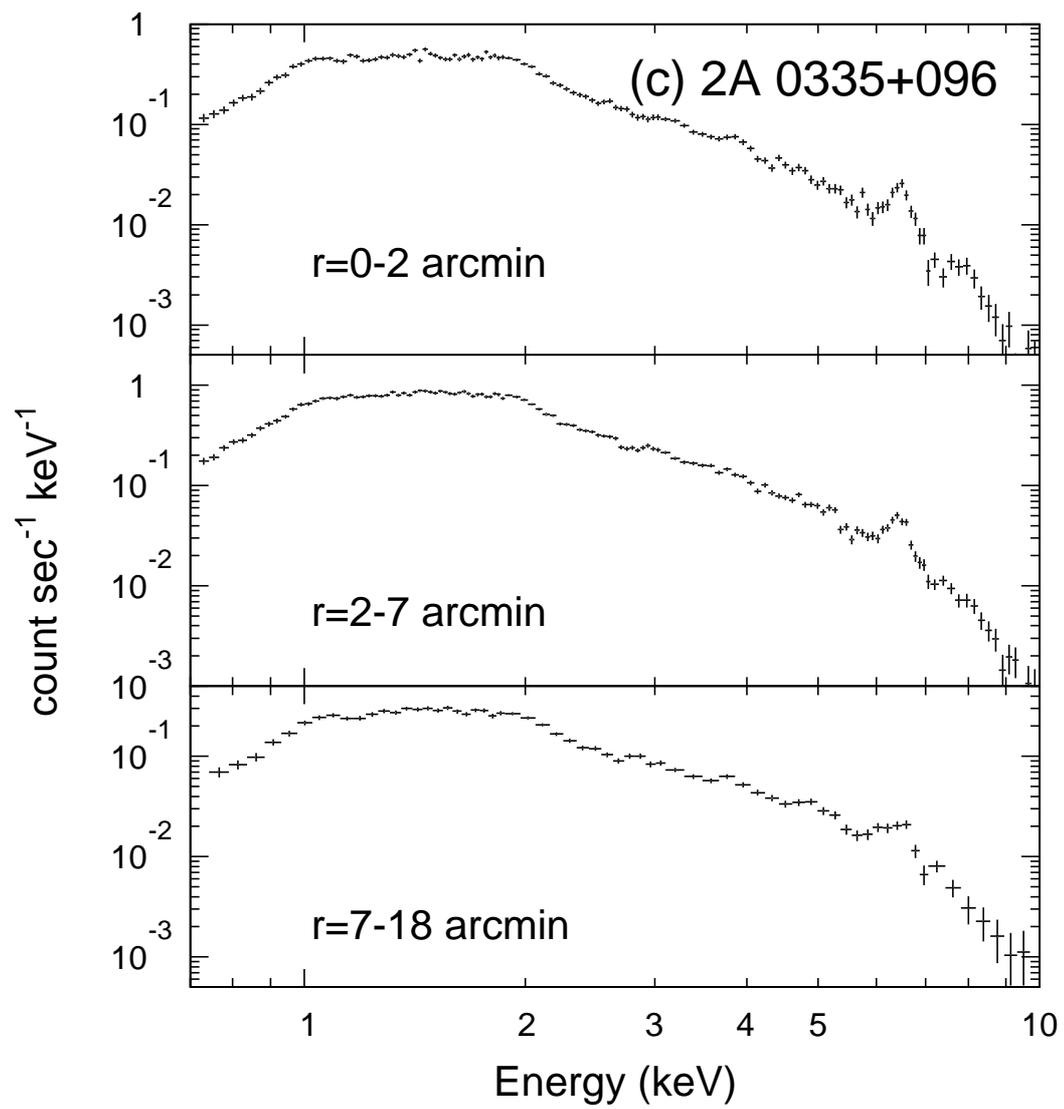
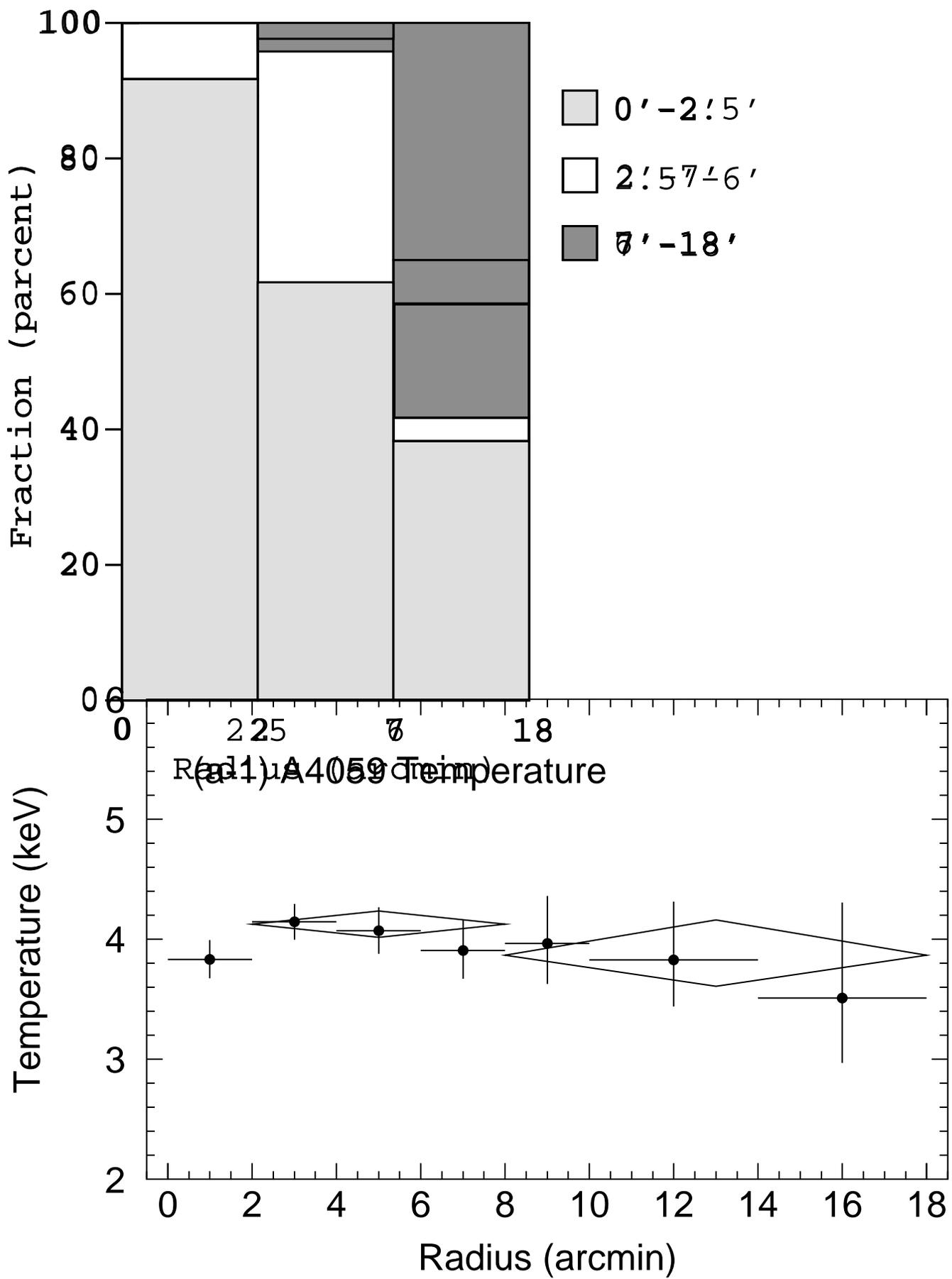
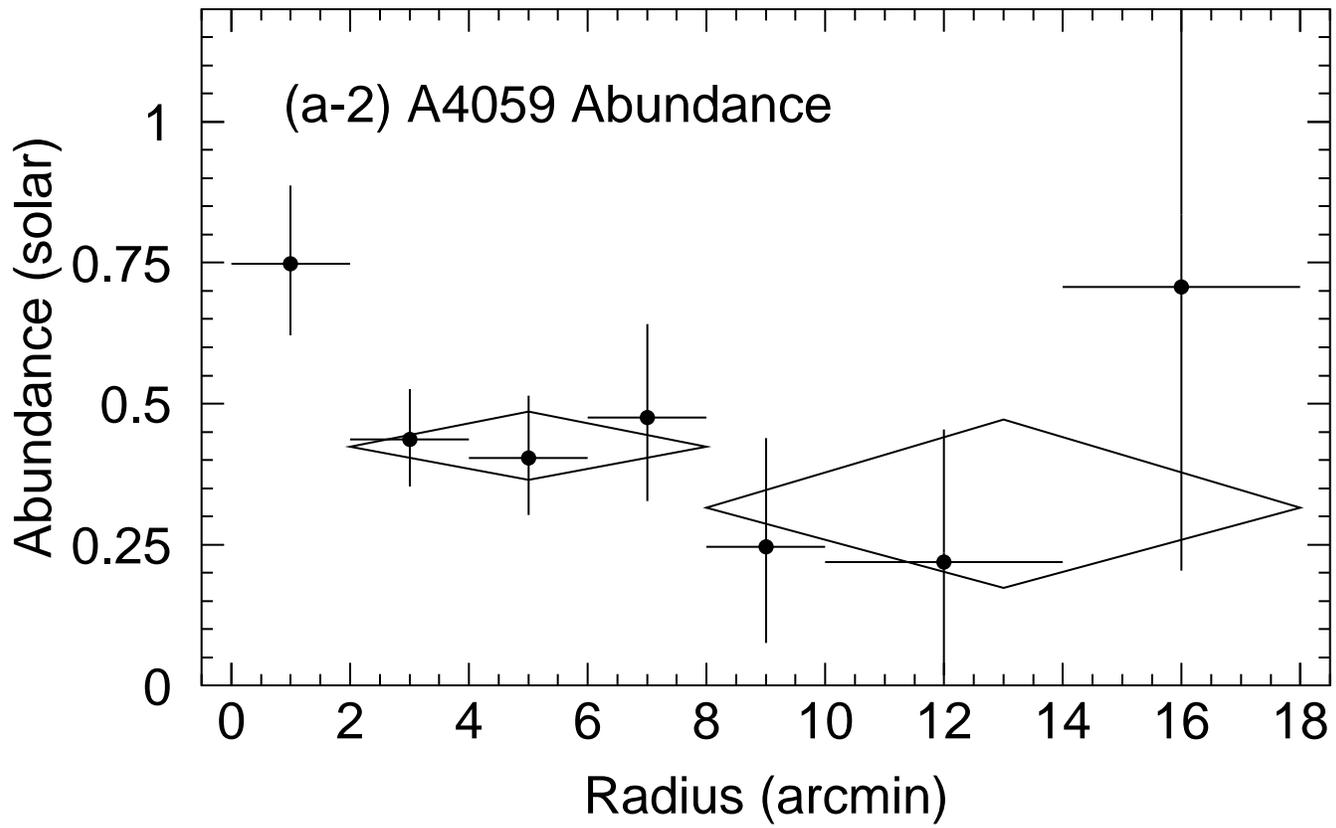
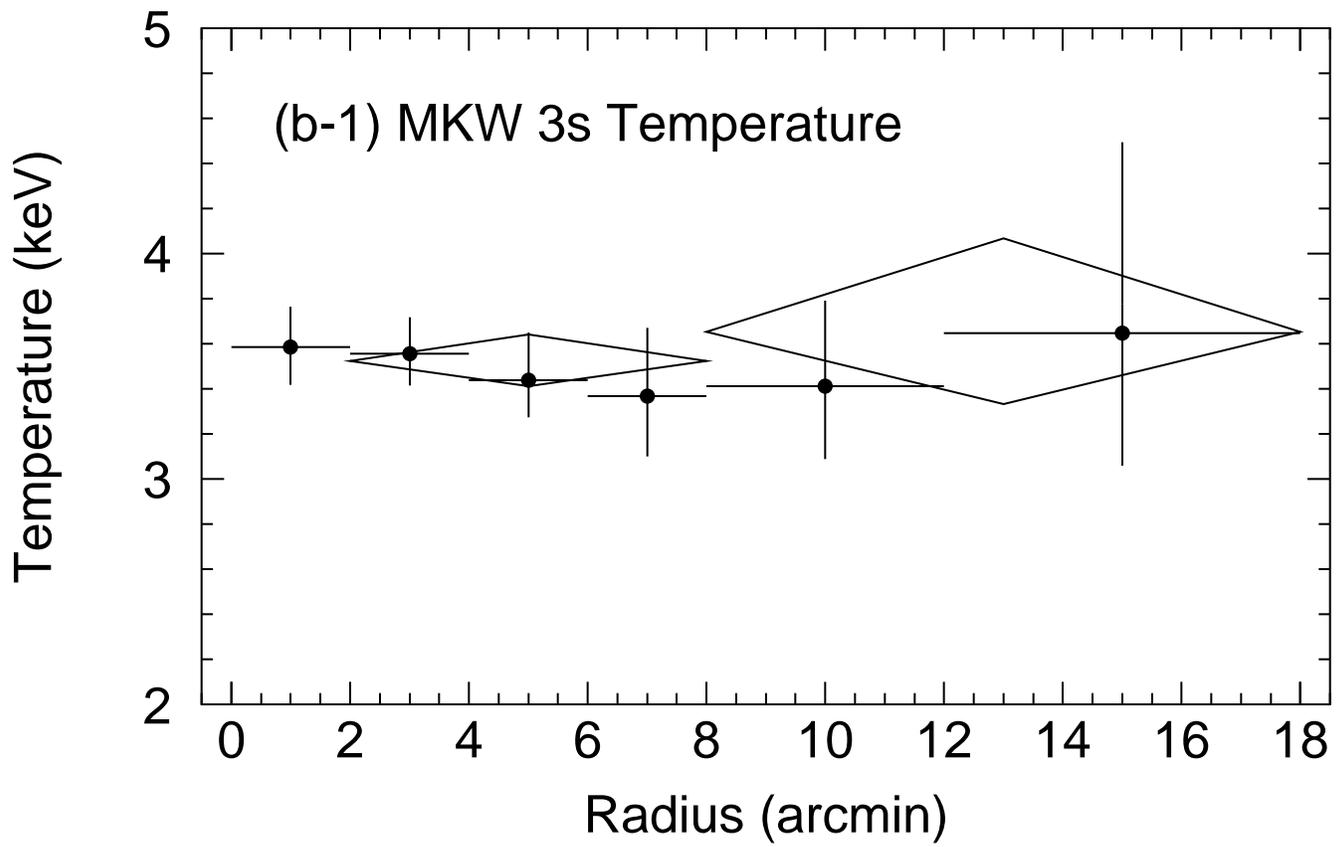
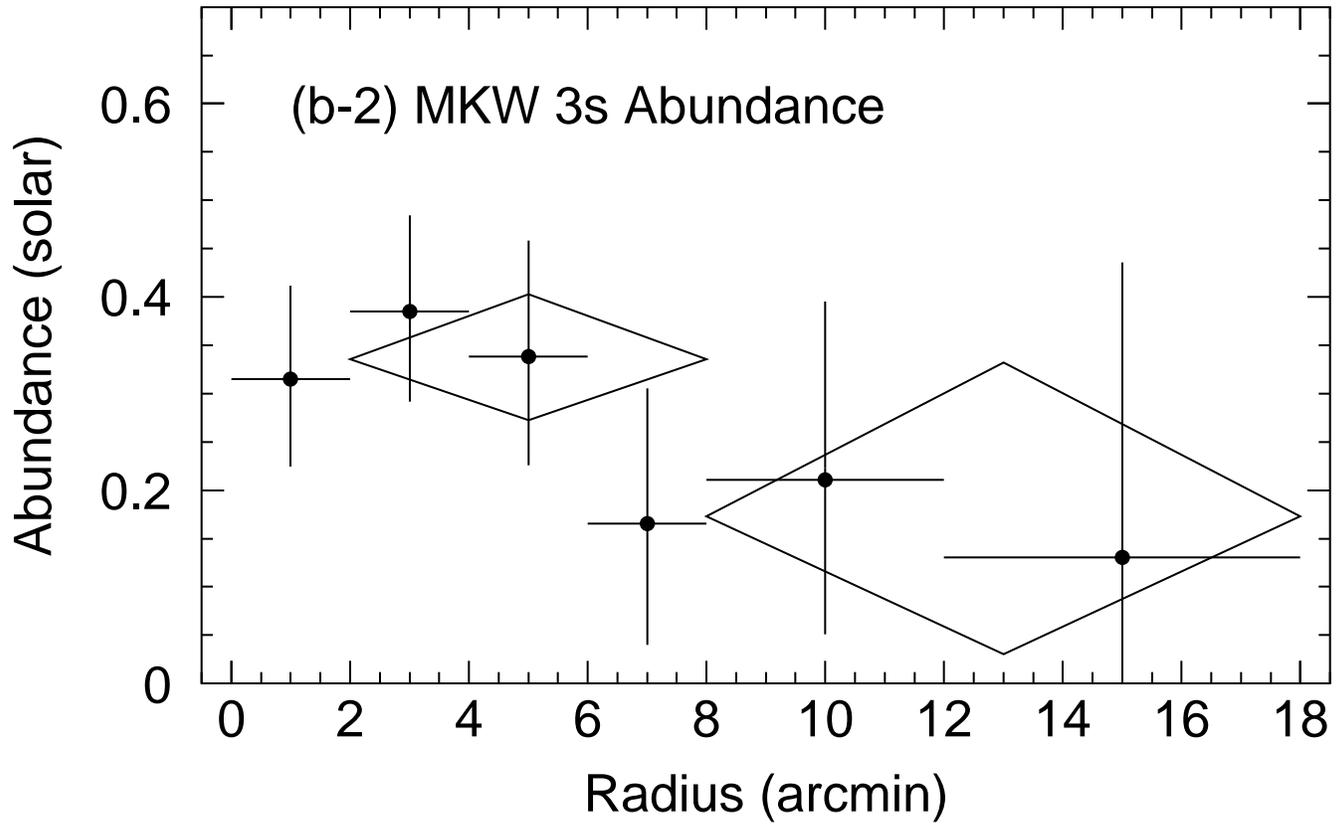
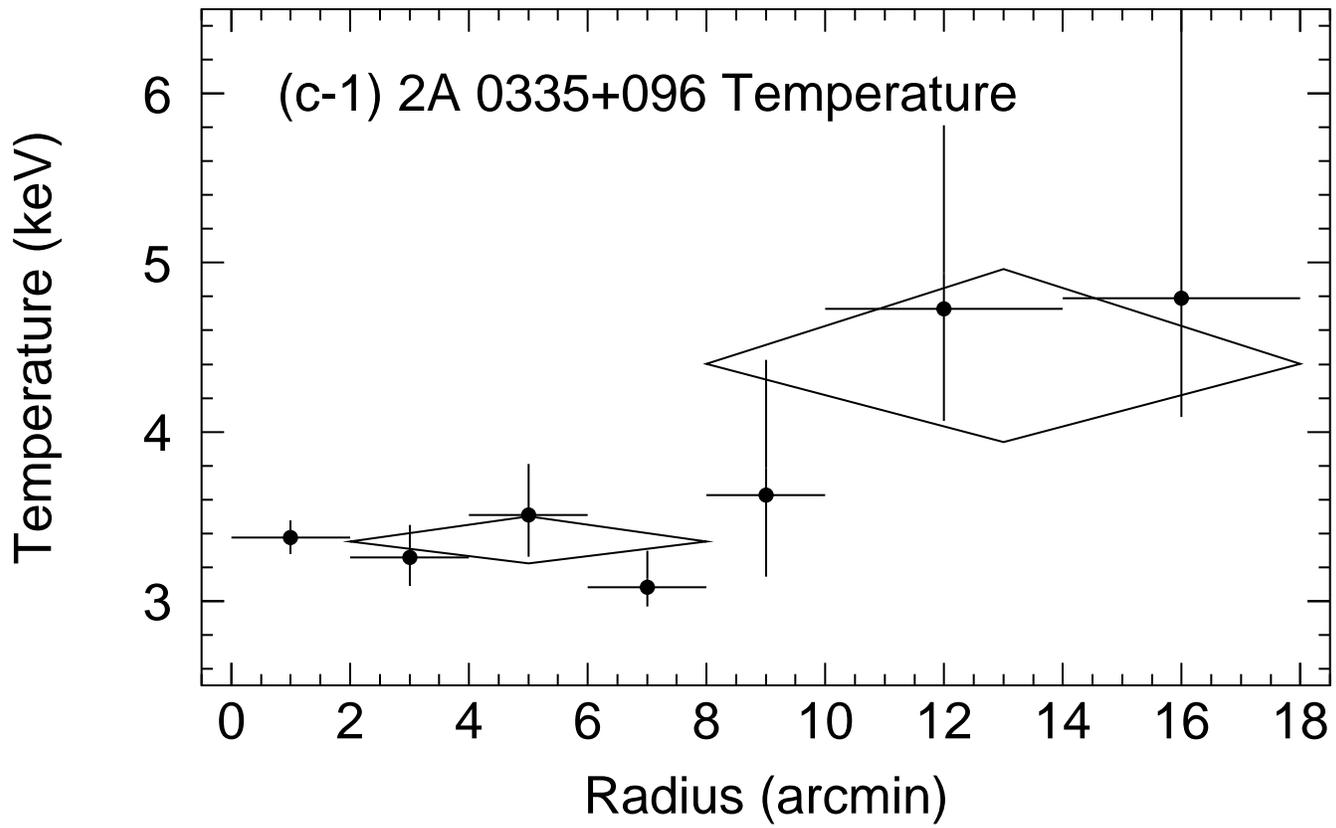
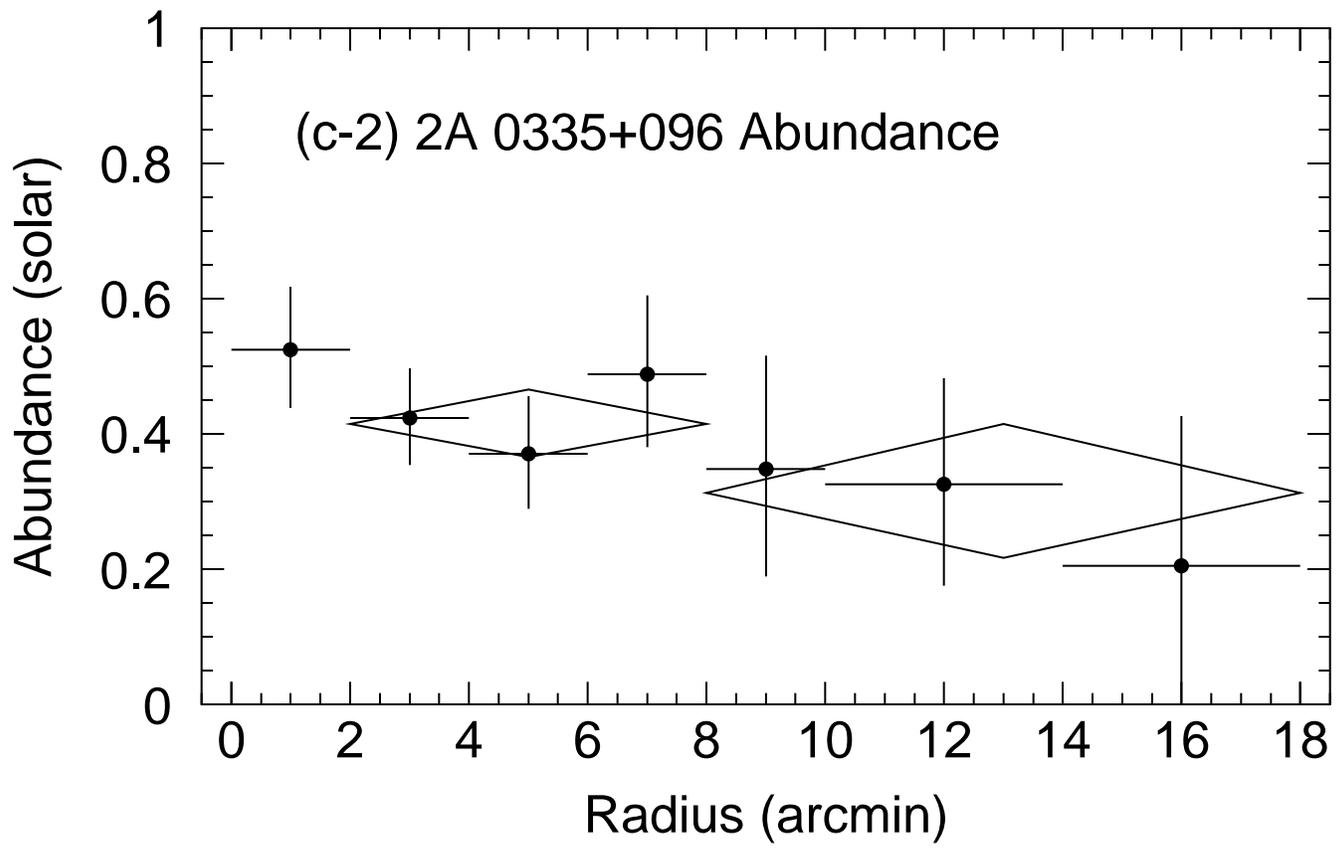
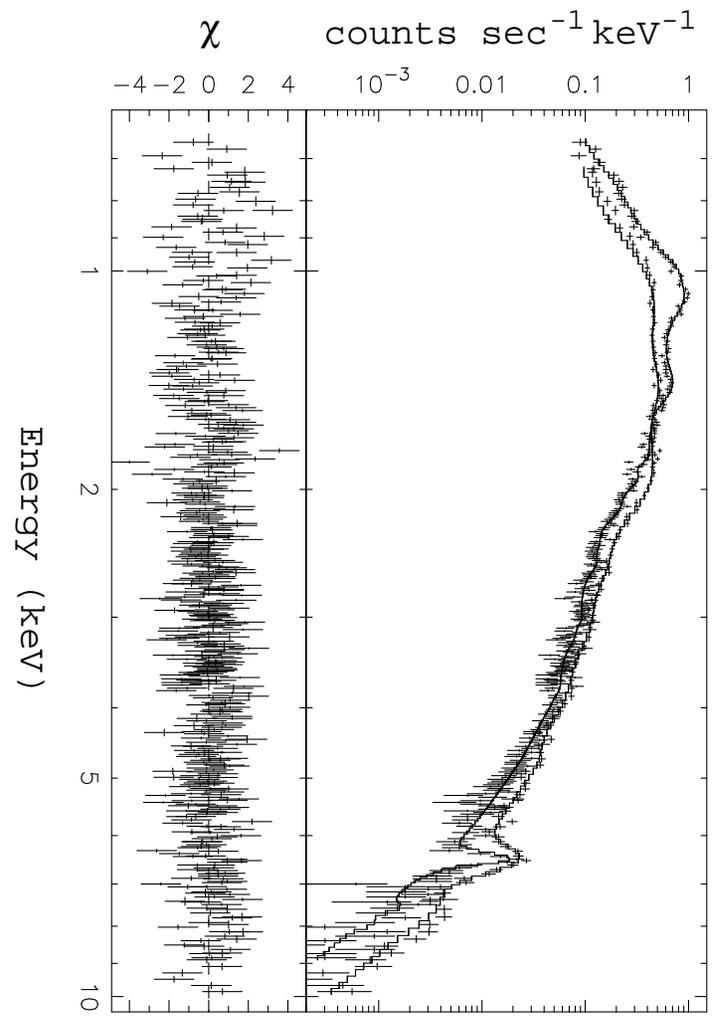
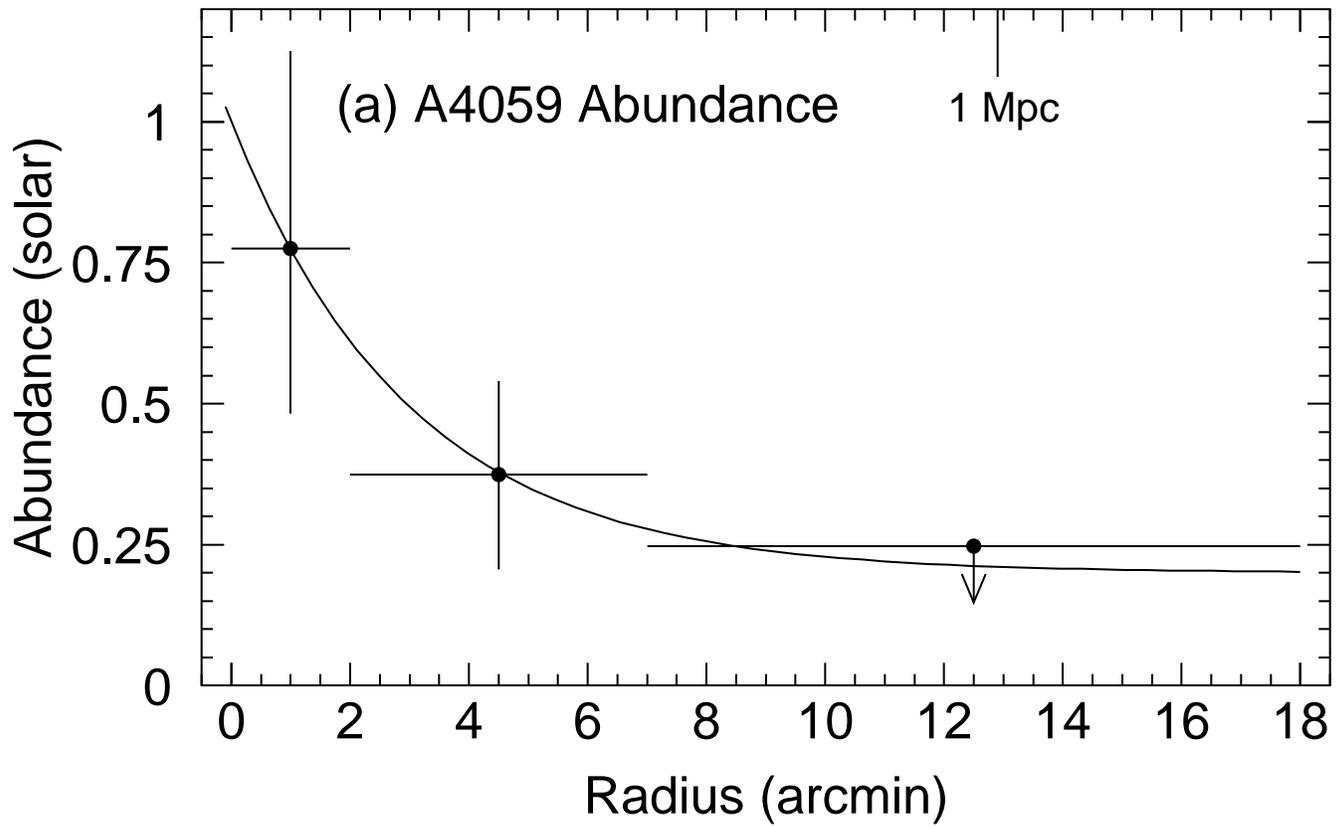
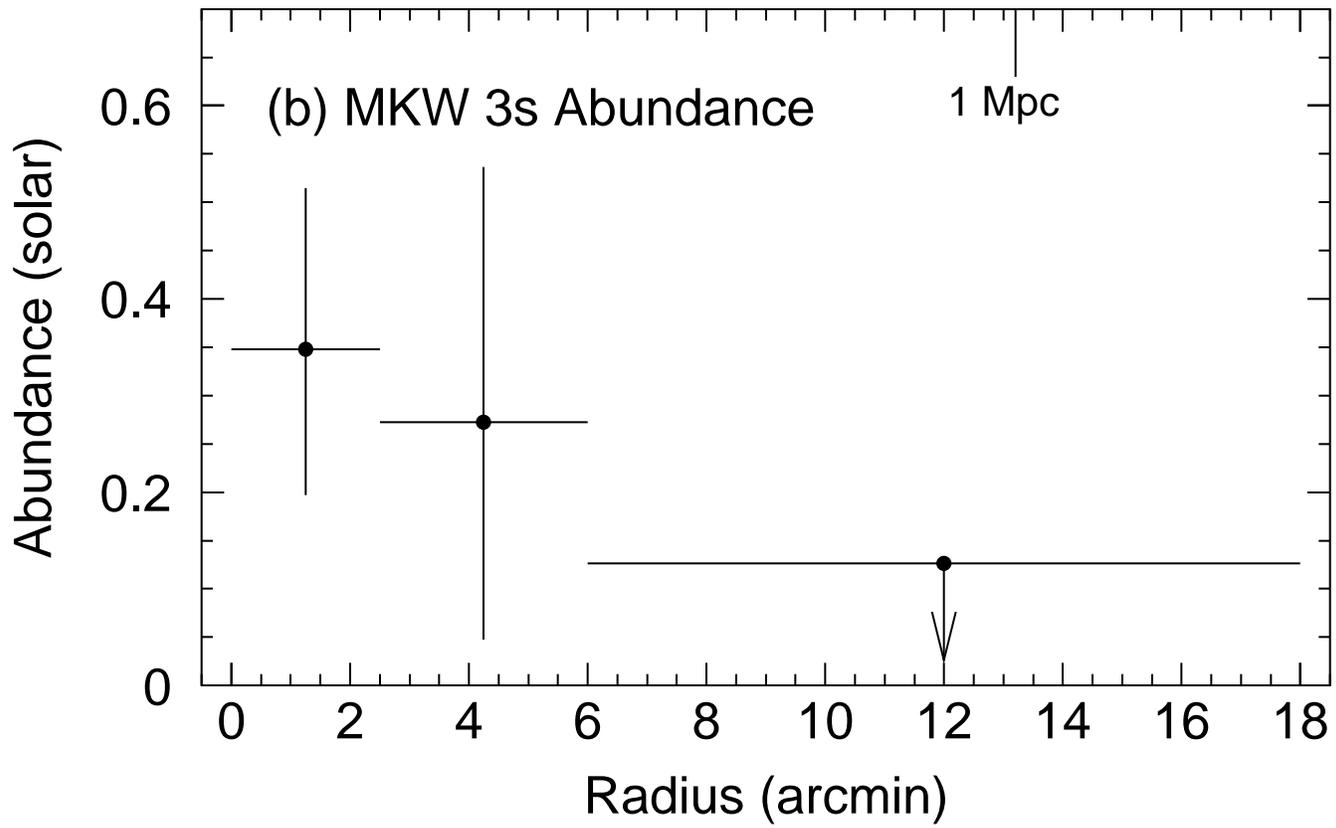
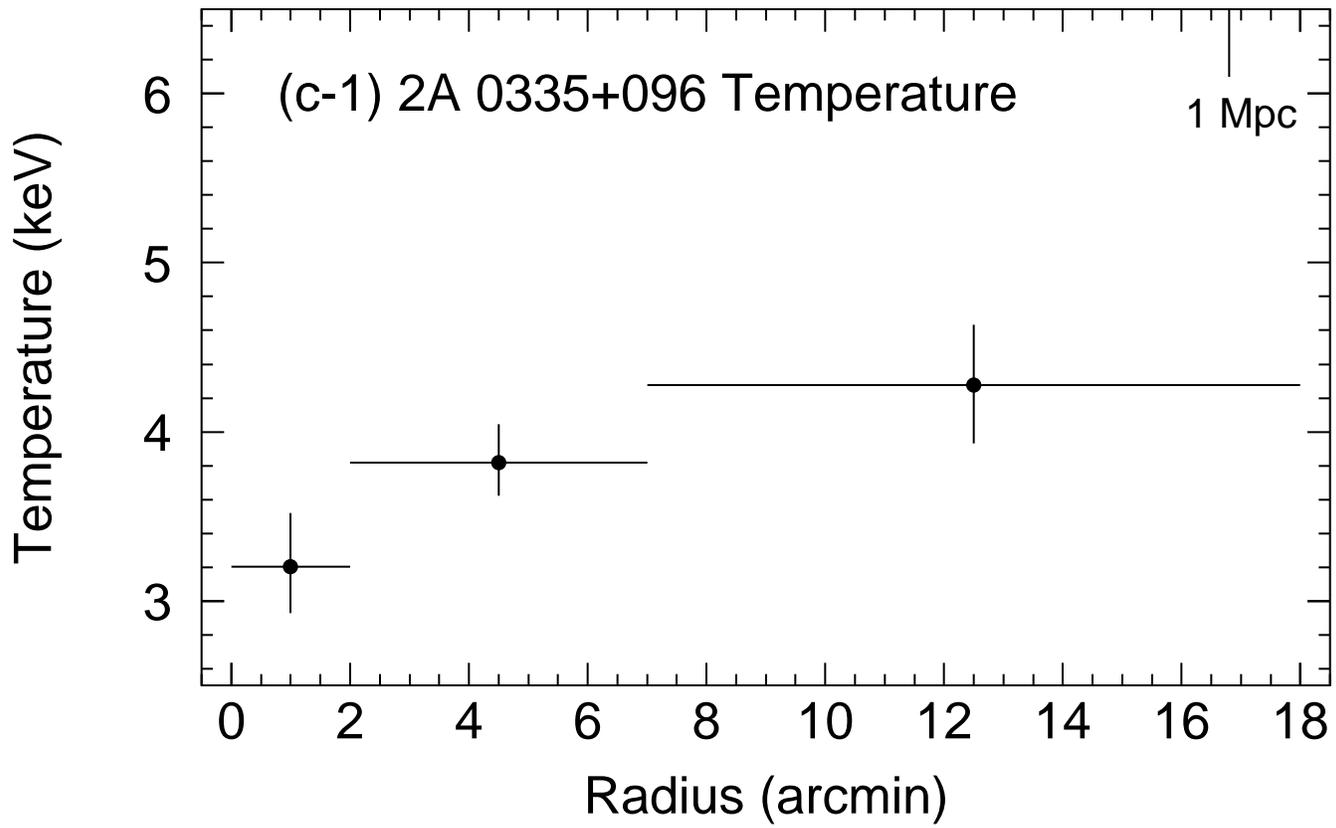
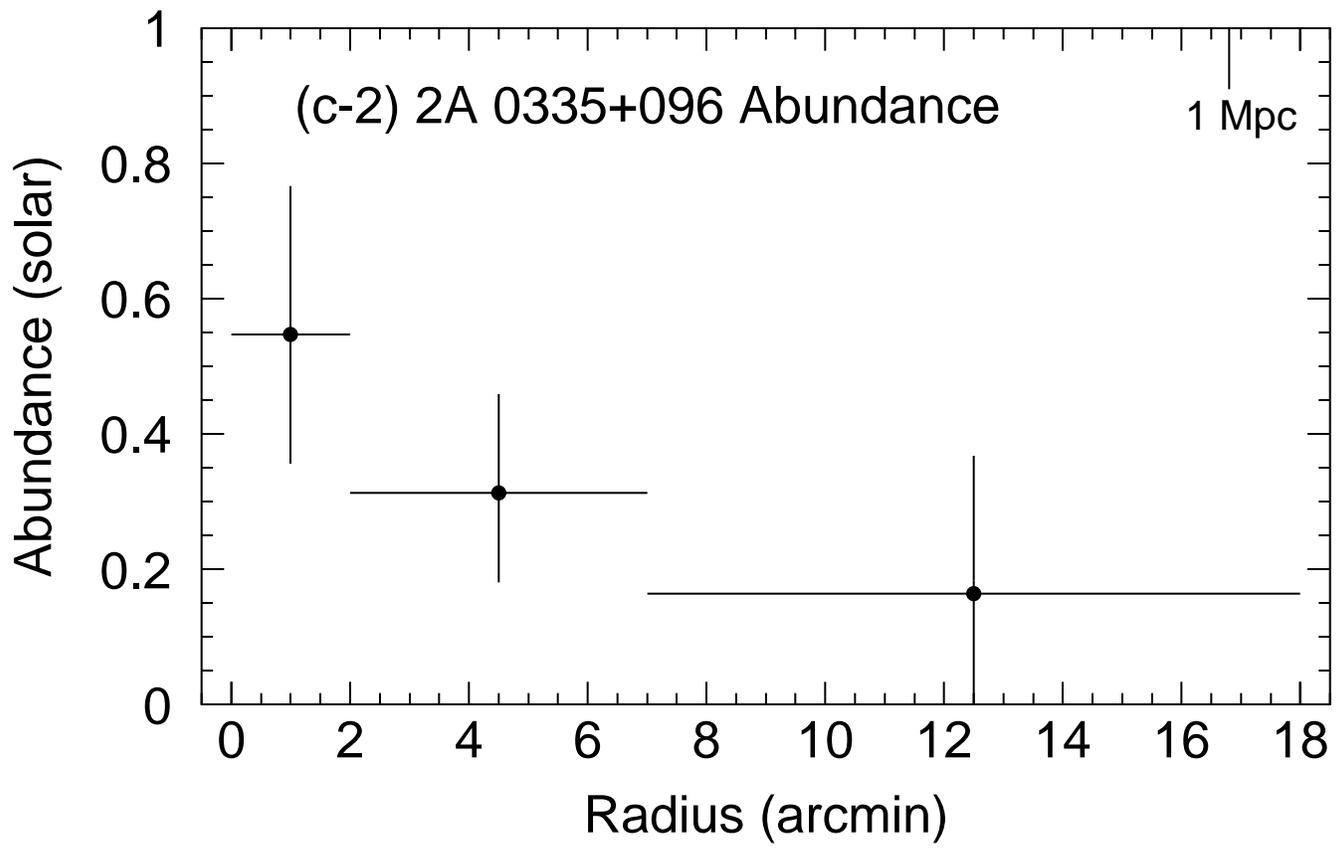
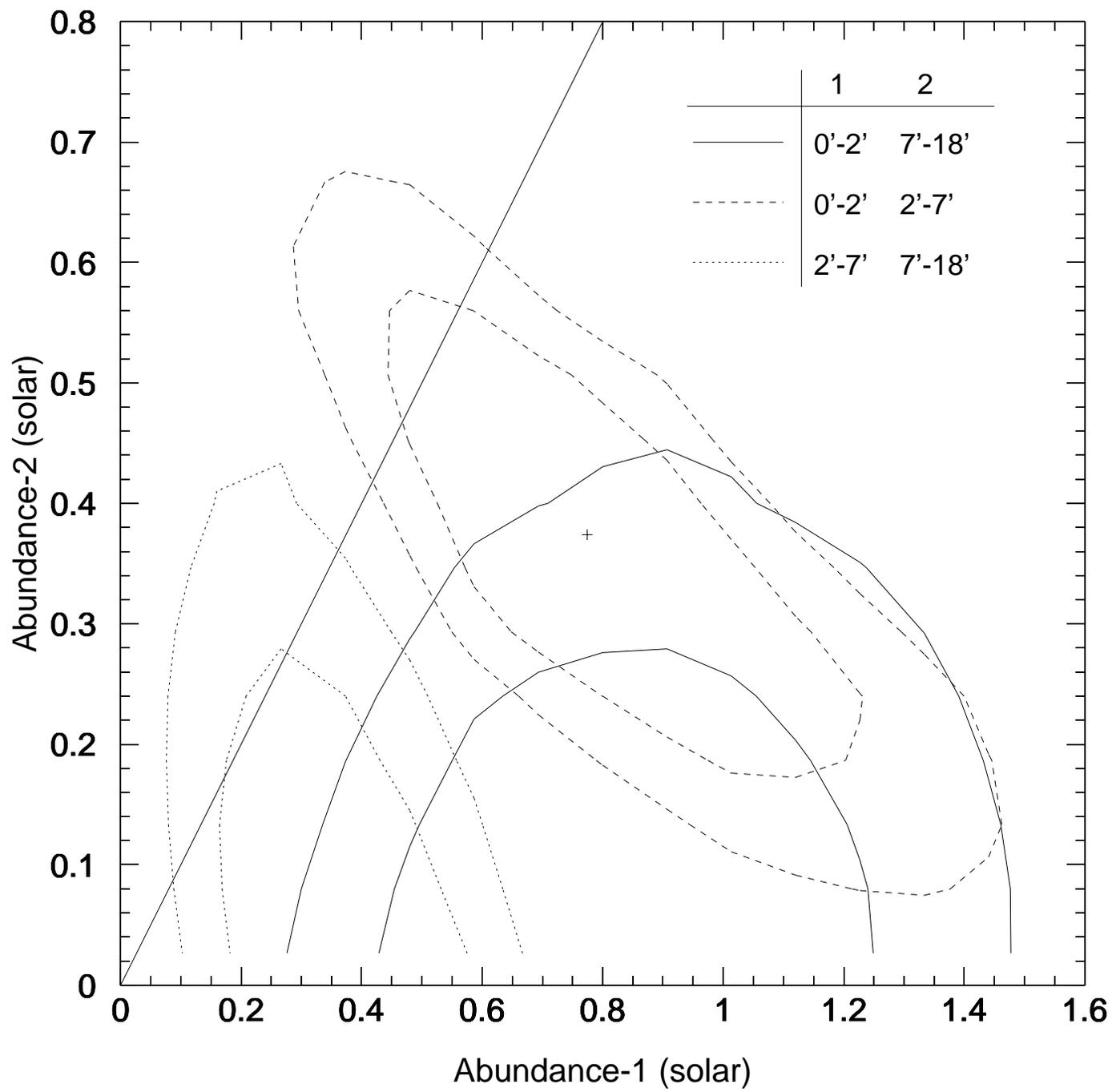
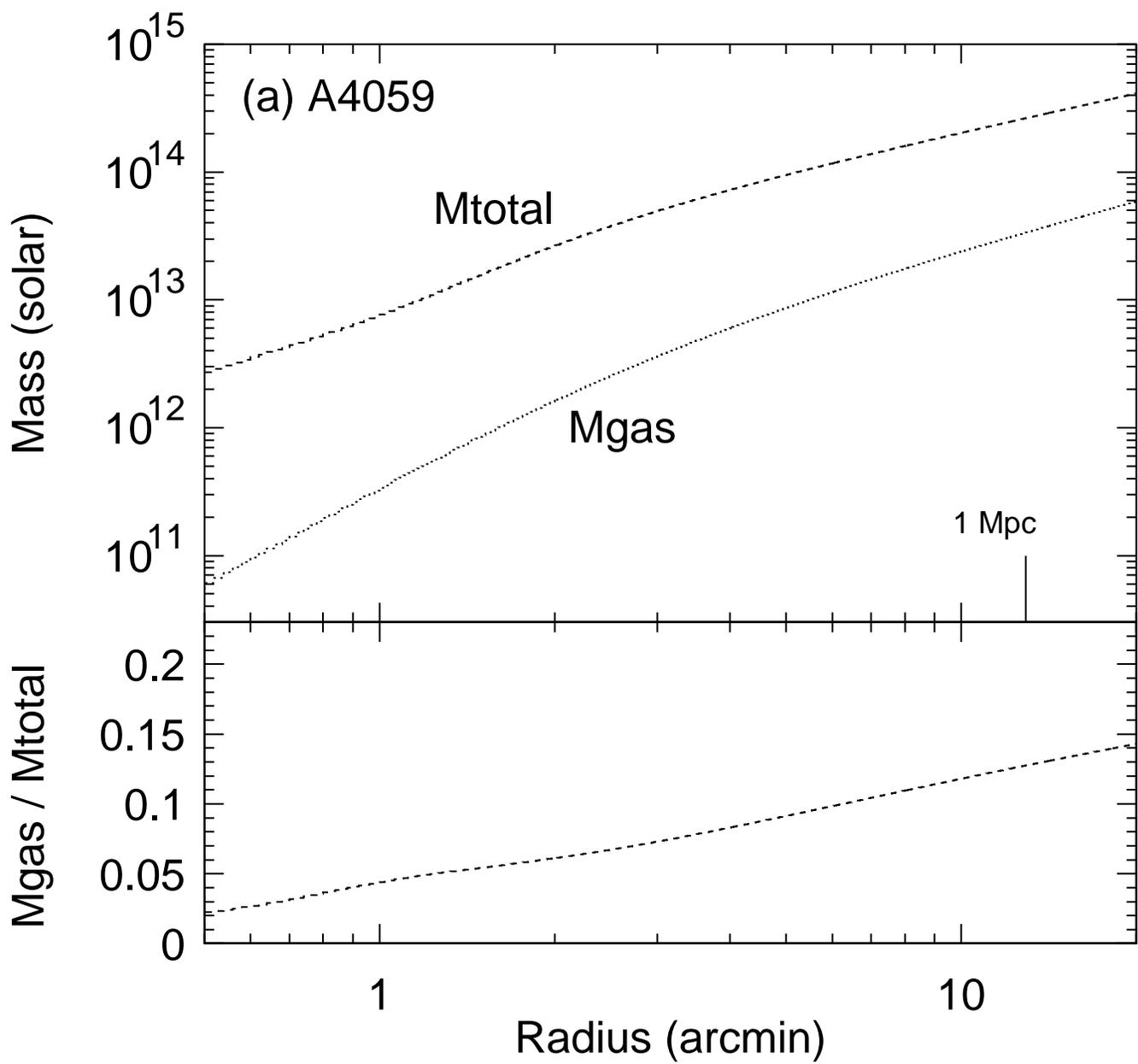
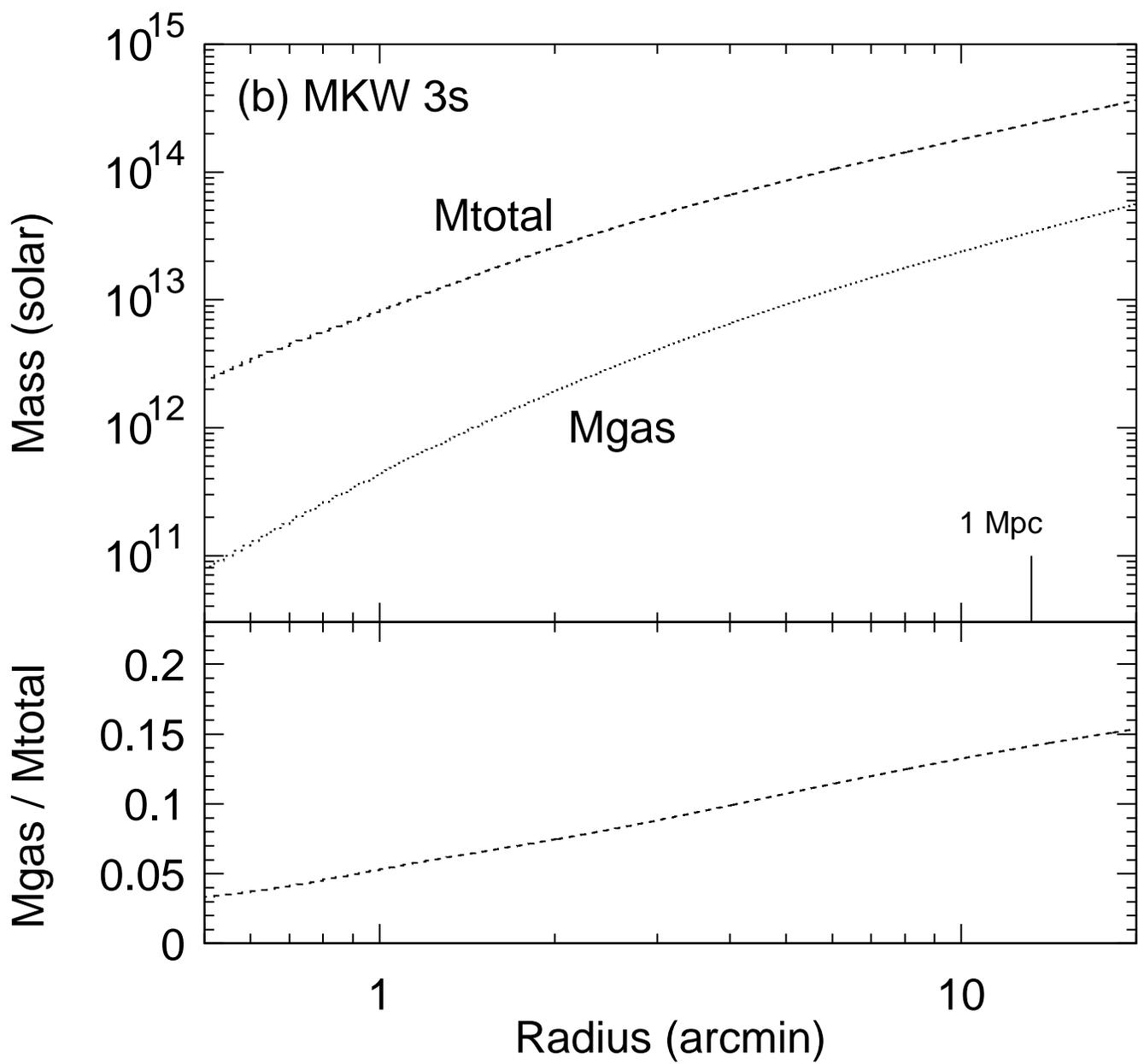
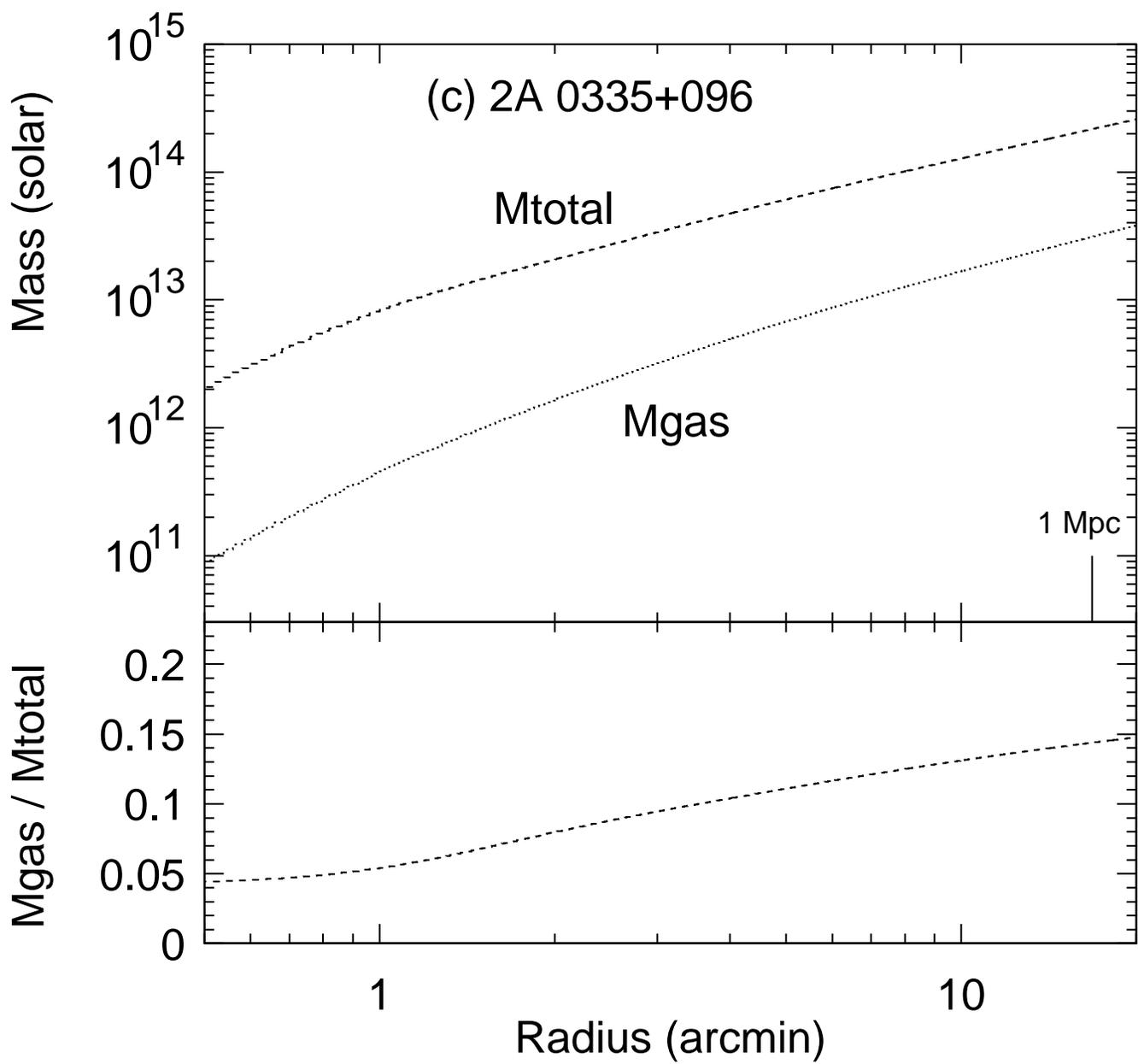
\end{document}